\documentclass[useAMS,usenatbib]{mn2e}
\usepackage{epsfig}
\newcommand{\mnras}{MNRAS}
\newcommand{\apj}{ApJ}
\newcommand{\apjs}{ApJS}
\newcommand{\apjl}{ApJL}
\newcommand{\aap}{A\&A}
\newcommand{\aj}{AJ}
\newcommand{\araa}{ARA\&A}
\newcommand{\nat}{Nature}
\newcommand{\be}{\begin{equation}}
\newcommand{\ee}{\end{equation}}
\newcommand{\beq}{\begin{eqnarray}}
\newcommand{\eeq}{\end{eqnarray}}
\newcommand{\atl}{ATLAS$^{\rm 3D}$}

\title[Fast \& Slow Rotators]
{The \atl\ Project-- VIII: Modelling the Formation and Evolution of Fast and Slow Rotator Early-Type Galaxies within $\Lambda$CDM }

\author[S.~Khochfar et al.]
{Sadegh Khochfar$^1$\thanks{E-mail: sadeghk@mpe.mpg.de},
Eric Emsellem$^{2,3}$, 
Paolo Serra$^{4}$,
Maxime Bois$^{2,3}$,
Katherine Alatalo$^5$,\newauthor
R.~Bacon$^3$,
Leo Blitz$^5$,
Fr\'ed\'eric Bournaud$^{6}$,
M.~Bureau$^{7}$,
Michele Cappellari$^{7}$,\newauthor
Roger L. Davies$^{7}$,
Timothy A. Davis$^{7}$,
P. T. de Zeeuw$^{2,8}$,
Pierre-Alain Duc,$^{9}$, \newauthor
Davor Krajnovi\'c$^2$,
Harald Kuntschner$^{10}$,
Pierre-Yves Lablanche$^{3}$,\newauthor
Richard M. McDermid $^{11}$,
Raffaella Morganti$^{4,12}$,
Thorsten Naab$^{13}$,
Tom Oosterloo$^{4,6}$,\newauthor
Marc Sarzi$^{14}$,
Nicholas Scott$^{7}$,
Anne-Marie Weijmans$^{15}$\thanks{Dunlap Fellow}
and Lisa M. Young$^{16}$\\
$^{1}$Max-Planck Institut f\"ur extraterrestrische Physik, PO Box 1312, D-85478 Garching, Germany\\
$^2$European Southern Observatory, Karl-Schwarzschild-Str. 2, 85748 Garching, Germany\\
$^3$Universit\'e Lyon 1, Observatoire de Lyon, Centre de Recherche Astrophysique de Lyon\\ and Ecole Normale Sup\'erieure de Lyon, 9 avenue Charles Andr\'e, F-69230 Saint-Genis Laval, France\\
$^{4}$Netherlands Institute for Radio Astronomy (ASTRON), Postbus 2, 7990 AA Dwingeloo, The Netherlands\\
$^5$Department of Astronomy, Campbell Hall, University of California, Berkeley, CA 94720, USA\\
$^6$Laboratoire AIM Paris-Saclay, CEA/IRFU/SAp Ð CNRS Ð Universit\'e Paris Diderot, 91191 Gif-sur-Yvette Cedex, France\\
$^{7}$Sub-department of Astrophysics, University of Oxford, Denys Wilkinson Building, Keble Road, Oxford OX1 3RH\\
$^{8}$Sterrewacht Leiden, Leiden University, Postbus 9513, 2300 RA Leiden, the Netherlands\\
$^{9}$Laboratoire AIM, CEA/DSM-CNRS-UniversitŽ Paris Diderot, Dapnia/Service d'Astrophysique, CEA-Saclay, 91191 Gif-sur-Yvette Cedex, France\\
$^{10}$Space Telescope European Coordinating Facility, European Southern Observatory, Karl-Schwarzschild-Str. 2, 85748 Garching, Germany\\
$^{11}$Gemini Observatory, Northern Operations Centre, 670 N. A`ohoku Place, Hilo, HI 96720, USA\\
$^{12}$Kapteyn Astronomical Institute, University of Groningen, Postbus 800, 9700 AV Groningen, The Netherlands\\
$^{13}$Max-Planck Institut f\"ur Astrophysik, Karl-Schwarzschild-Str. 1, 85741 Garching, Germany\\
$^{14}$Centre for Astrophysics Research, University of Hertfordshire, Hatfield, Herts AL1 9AB, UK\\
$^{15}$Dunlap Institute for Astronomy \& Astrophysics, University of Toronto, 50 St. George Street, Toronto, ON M5S 3H4, Canada\\
$^{16}$Physics Department, New Mexico Institute of Mining and Technology, Socorro, NM 87801, USA\\
}

\date{Draft \today }

\pagerange{\pageref{firstpage}--\pageref{lastpage}} \pubyear{2011}

\def\simpropto{\lower.2ex\hbox{$\; \buildrel \propto \over \sim \;$}}
\def\ltsim{\lower.5ex\hbox{$\; \buildrel < \over \sim \;$}}
\def\gtsim{\lower.5ex\hbox{$\; \buildrel > \over \sim \;$}}

\begin{document}

\label{firstpage}

\maketitle

\clearpage
\begin{abstract}
We propose a simple model for the origin of fast and slow rotator early-type galaxies (ETG) within the hierarchical $\Lambda$CDM scenario, that is based on the assumption that the mass fraction of stellar discs in ETGs is a proxy for the specific angular momentum expressed via $\lambda_R$.  Within our model we reproduce the fraction of fast and slow rotators as a function of magnitude in the \atl survey, assuming that fast rotating ETGs have at least $10\%$ of their total stellar mass in a disc component. In agreement with \atl observations we find that slow rotators are predominantly galaxies with $ M_* > 10^{10.5}$ M$_{\odot}$ contributing $\sim 20 \%$ to the overall ETG population.  We show in detail that the growth histories of fast and slow rotators are  different, supporting the classification of ETGs into these two categories. Slow rotators accrete between $\sim 50 \% -90\%$ of their  stellar mass from  satellites and their most massive progenitors have on average up to 3 major mergers during their evolution. Fast rotators in contrast, accrete less than $50\%$ and have on average less than one major merger in their past. 

We find that the underlying physical reason for the different growth histories is the slowing down and ultimately complete shut-down of gas cooling in massive galaxies. Once cooling and associated star formation in disc stops, galaxies grow via infall from satellites. Frequent minor mergers thereby, destroy existing stellar discs via violent relaxation and also tend to lower the specific angular momentum of the main stellar body, lowering $\lambda_R$ into the slow rotator regime. 

On average the last gas-rich major merger interaction in slow rotators happens at $z > 1.5$, followed by a series of minor mergers. These results support the idea that kinematically decoupled cores (KDC) form during gas-rich major mergers at high-z followed by minor mergers, which build-up the outer layers of the remnant, and make remnants that are initially too flat compared to observations become rounder. Fast rotators are less likely to form such KDCs due to the fact that they have on average less than one major merger in their past.

Fast rotators in our model have different formation paths. The majority, $78 \%$  has bulge-to-total stellar mass ratios $B/T> 0.5$ and managed to grow stellar discs due to continued gas cooling or bulges due to frequent minor mergers. The remaining $22 \%$ live in high density environments and consist of low $B/T$ galaxies with gas fractions below $15 \%$, that have exhausted their cold gas reservoir and have no hot halo from which gas can cool. These fast rotators most likely resemble the flattened disc-like fast rotators in the \atl survey.  
  
Our results predict that ETGs can change their state from fast to slow rotator and vice versa, while the former is taking place predominantly at low z $(z < 2)$, the latter is occurring during cosmic epochs when cooling times are short and galaxies gas-rich. We predict that the ratio of the number density of slow to fast rotators is a strong function of redshift, with  massive $( > 10^{10}$ M$_{\odot}$) fast rotators being more than one order of magnitude more frequent at $z \sim 2$.

\end{abstract}


\section{Introduction}
Much attention has been paid to the modelling of the formation of early-type galaxies (ETG) over the last three decades 
\citep[e.g.][]{1972ApJ...178..623T,1978MNRAS.184..185W,1983MNRAS.205.1009N,1992ARA&A..30..705B,1996ApJ...464..641M,2003ApJ...597..893N,2004A&A...418L..27B,2006ApJ...650..791C,2010ApJ...723..818H}. Once believed to be coeval featureless stellar systems in virial equilibrium that formed from a 'monolithic' collapse \citep{1974MNRAS.166..585L}, an  ever-increasing wealth of observational data has revealed a multitude of distinct physical properties: in their rotational support 
\citep{1978MNRAS.183..501B,1983ApJ...266...41D}, isophotal shape \citep{1989A&A...217...35B,1996ApJ...464L.119K}, photometric profiles \citep{1994AJ....108.1598F,1995AJ....110.2622L,2009ApJS..182..216K}, and in their stellar angular momentum \citep{2007MNRAS.379..401E,2007MNRAS.379..418C}.  Driven by such observations theoretical models have been put to the test, and in particular the $\Lambda$CDM paradigm for structure formation itself  within which structure grows hierarchically via a sequence of mergers and accretion events \citep{1978MNRAS.183..341W}.  

Early numerical N-body simulations \citep{1972ApJ...178..623T} identified galaxy mergers as a viable mechanism to transform dynamical cold stellar systems into dynamical hot ones resembling early-type galaxies with a de Vaucouleurs-like  light profile \citep[see also][]{1992ARA&A..30..705B,2006MNRAS.369..625N}. The main driver for this morphological transformation being violent relaxation \citep{1967MNRAS.136..101L} during which stars get redistributed. 
The merger scenario subsequently has been tested against two main categories of observational data: global properties in large homogenous surveys \citep[e.g.][]{2003AJ....125.1817B}  and detailed high-resolution observations of an inhomogeneous set of individual galaxies \citep[e.g.][]{2002MNRAS.329..513D}. 

Galaxy formation models, in particular semi-analytic models,  employing the merger scenario, reproduce successfully the mass function, colour distribution, star formation history  and metallicity of early-type galaxies as observed e.g. in the Sloan Digital Sky Survey (SDSS) \citep[e.g.]{2006MNRAS.366..499D}. However, they still have problems in e.g. recovering the trend in $\alpha/Fe$ with stellar velocity dispersion \citep{2005ApJ...621..673T,2005MNRAS.363L..31N}  or the joint distribution in sizes and velocity dispersion at $z=0$ \citep{2009ApJ...698.1232V,2010MNRAS.405..948S}. The latter has also been reflected in mismatches to the fundamental plane  of early type galaxies \citep{2007MNRAS.376.1711A}. While the deviations regarding the sizes and velocity dispersions likely stem from too efficient in-situ star formation in progenitors of present-day early-type galaxies \citep{2010MNRAS.405..948S}, the failure to produce enough $\alpha$-elements is most likely  connected to a combination of too much residual star formation \citep{2005ApJ...621..673T} and satellite infall \citep{ks06}.           

The inner parts of early-types show further detailed characteristics in the form of steep increasing  power-law or  core like profiles \citep{1994AJ....108.1598F,1995AJ....110.2622L,2009ApJS..182..216K}. The former has been associated with cold gas in discs that loses its angular momentum during the merging process \citep[e.g.][]{1991ApJ...370L..65B}  and settles to the centre of the remnant where it triggers a star burst, thereby steepening the light profile \citep{2008ApJ...679..156H}.  Core galaxies on the other hand are most likely the product of a {\it non-dissipative} merger event between two galaxies hosting super-massive black holes (SMBH) at their centre \citep{2001ApJ...563...34M,2005MNRAS.359.1379K,2009ApJS..181..486H}. During the inward spiralling the SMBH pair kick out stars via three-body interaction and produces a density core in the centre of the remnant \citep{2001ApJ...563...34M}.   

Classifying elliptical galaxies by the deviation of their isophotal shape from pure ellipses has been suggested as a method to separate the population of elliptical galaxies into two classes with box-like (boxy) and disk-like deviations (disky) \citep{1989A&A...217...35B}. Numerical simulations show that equal mass mergers in general produce remnants with boxy isophotes due to  stronger gravitational interaction and associated violent relaxation \citep{2000MNRAS.316..315B,2003ApJ...597..893N} efficiently destroying progenitor discs. In contrast unequal mass major mergers  with 
$2 \le M_1/M_2 \le 4$ result on average in disky remnants for the majority of projections under which they are viewed \citep{2003ApJ...597..893N}.
Observationally, the average isophotal shape of elliptical galaxies turns toward boxy going up in mass \citep[see howerver][]{pap3} indicating a transition in the formation process of massive ellipticals that cannot be attributed to the mass ratio during the last major merger alone, which in general does not change significantly enough in frequency as a function of mass  \citep{2005MNRAS.359.1379K}. In addition unequal mass mergers of early-type galaxies are able to transform disky ellipticals into boxy ones in the majority of simulated cases \citep{2006ApJ...636L..81N}. Within a cosmological context the relative fraction of dry, early-type major mergers increases with stellar mass of the remnant \citep{2003ApJ...597L.117K} providing the additional channel needed to recover the observed trend with mass \citep{2005MNRAS.359.1379K}. In a detailed study of the orbital content of stars in merger remnants \citet{2005MNRAS.360.1185J} showed that the individual mixture of orbits is the deciding factor on the isophotal shape of disky and boxy elliptical galaxies.  

Making full use of the dynamical information available in integral-field observations the SAURON sample \citep{2002MNRAS.329..513D} measured the specific baryonic angular momentum content via the proxy $\lambda_R$ within one $R_e$ of a representative sample of early-type galaxies and separate them into fast and slow rotators according to their $\lambda_R$ value, in a robust way that is nearly insensitive to projection effects  \citep{2007MNRAS.379..401E,2007MNRAS.379..418C}. Observationally, fast rotators present regular rotation patterns aligned with the photometry, while slow rotators have low angular momentum content and show misalignments between the photometry and the velocity axes and often exhibit kinematically distinct cores \citep{2008MNRAS.390...93K}. The division into fast and slow rotators is further supported by recent simulations of \citep[hereafter Paper VI]{pap6}
that show that fast (resp. slow) rotators have a high (resp. low) angular momentum content, their photometric and kinemetric position angle are aligned (resp. misaligned), they present (resp. do not present) regular velocity patterns.
While the SAURON survey was biased towards massive ETGs \citep{2002MNRAS.329..513D} the \atl survey \citep[][hereafter Paper I]{pap1} for the first time allows to give a complete view of the local ETG population within a radius of $42 $ Mpc using the full information of integral-field observations. The combination of completeness with detailed structural information and complementary observations on e.g. the HI \citep{paolo} or molecular  \citep{pap4}  gas masses, make the \atl survey an ideal data set to investigate the formation of ETGs. 

In this paper we attempt to model fast and slow rotators within the standard $\Lambda$CDM paradigm and analyse specific differences in their formation. The structure of the paper is as follows: we start by summarising the basic model ingredients used in our semi-analytic modelling (SAM) approach  in section \ref{mod}, followed by showing the performance of the model in section \ref{modnor}. In section \ref{fas} we introduce our model for fast and slow rotators and investigate the robustness of our assumptions before we analyse in detail differences in their formation in section \ref{demo} and predict their redshift evolution in section \ref{z}.  Section \ref{var} is devoted to the question of cosmic variance in the \atl sample followed by sections \ref{sum} \& \ref{con} in which we summarise our results and conclude respectively.
 
\section{The modelling approach} \label{mod}
We apply the semi-analytic modelling (SAM) approach to galaxy formation pioneered in a series of papers \citep[e.g.][]{1977ApJ...211..638S,1978MNRAS.183..341W,1991ApJ...379...52W}. The main strategy behind SAMs is to first calculate the collapse and merging history of individual dark matter halos,  and secondly 
to approximate the more complex physics of the baryons inside 
these dark matter halos by analytic models \citep[see e.g.][]{1999MNRAS.303..188K,1999MNRAS.310.1087S,2000MNRAS.319..168C,2003MNRAS.343...75H,2006MNRAS.370.1651C,2006MNRAS.365...11C,2006MNRAS.370..645B,2006ApJ...648..820K,2007MNRAS.375....2D,2009ApJ...700L..21K}. Each of the dark matter halos will consist of three 
main components which are distributed among individual galaxies inside  
them: a stellar, a cold, and  a hot gas component, where the latter  is
only attributed to {\it central} galaxies, which are the most massive galaxies
inside individual halos and are typically observed to reside in extended X-ray 
emitting hot haloes of gas \citep{2010ApJ...715L...1M}.
In the following sections, we will describe briefly the recipes used to
 calculate these different components which 
are mainly based on recipes presented in e.g. 
\citet{1999MNRAS.303..188K} (hereafter, K99) and 
\citet{spr01} (hereafter, S01), and we refer readers for more details 
on the basic model implementations to their work and references therein. 
In the remainder of this paper we call SAM the model implementation 
presented in \citet{2003ApJ...597L.117K,2005MNRAS.359.1379K}  which is summarised in the following parts of this section. With our SAM we model a cosmic volume of $V_{A3D}=10^5$ Mpc$^3$, which is comparable to the volume of the \atl survey (Paper I). Throughout this paper, we use the following set of cosmological parameters consistent with 7-year WMAP data \citep{2010arXiv1001.4538K}:
$\Omega_0=0.26$, $\Omega_{\Lambda}=0.74$, $\Omega_b/\Omega_0=0.17$, $n_s=0.97$,  
$\sigma_8=0.8$ and $h=0.71$. 

\subsection{Dark Matter}
The mass function of dark matter haloes within our volume is calculated using the analytic fitting formula of 
\citet{2001MNRAS.321..372J}. In addition, to investigate the effects of cosmic variance (see Section \ref{var}) we have run a cosmological N-body simulation of a $(100$ Mpc$/h)^3$ volume with GADGET-2 \citep{2005MNRAS.364.1105S}. The particle resolution of the simulation is $10^9$ M$_{\odot}/h$, and we generate the $z=0$ halo mass function using the friends-of-friends algorithm. 
  
The corresponding halo merging histories are calculated using a Monte-Carlo approach based on the 
Extended-Press-Schechter formalism \citep[e.g.][]{1993MNRAS.262..627L} following the method 
presented in \citet{1999MNRAS.305....1S}. This approach has 
been shown to produce merging histories and progenitor distributions in 
reasonable agreement with results from N-body simulations of cold dark matter 
structure formation in a cosmological context \citep{2000MNRAS.316..479S}. 
The merging history of dark matter halos is reconstructed by breaking 
 each halo up into progenitors above a limiting minimum progenitor 
mass $M_{min}$. This mass cut needs 
to be chosen carefully as it ensures that the right galaxy population and 
merging histories are produced within the model. Progenitor halos
 with masses below $M_{min}$ are declared as {\it accretion} events and 
their histories are not followed further back in time. 
Progenitors labelled as accretion events should ideally not host any 
significant galaxies in them and be composed mainly of primordial gas. 
To achieve a good compromise between accuracy and computational time, we estimate $M_{min}$  by running several simulations with different resolutions 
and chose the resolution for which results in the galaxy mass range of 
interest are independent of the specific choice of $M_{min}$. 
Changing the mass resolution mainly affects our results at low galaxy mass 
scales, leaving massive galaxies, the focus of this work, literally unaffected.
Throughout this paper we will use $M_{min}= 10^{10}$ M$_{\odot}$  
which produces numerically stable results for galaxies with stellar masses 
greater $\sim 6 \times 10^{9}$ M$_{\odot}$, which corresponds to the mass limit in the \atl survey. For simplicity we assume in  the following that the dark matter profiles are that of truncated  isothermal spheres \citep{1999MNRAS.303..188K}.

\subsection{Baryonic Physics}
Once the merging history of the dark matter 
component has been calculated, it is
 possible to follow the evolution of the baryonic content in these halos 
forward in time. We assume each halo  consists of three components: 
hot gas ($M_{hot}$), cold gas ($M_{gas}$) and stars ($M_*$), where the latter two components can be distributed 
 among individual galaxies, labelled {\it central} and {\it satellite}, inside a single  dark matter halo. The hot gas component however, is only associated with central galaxies. The stellar component 
of each galaxy is additionally divided into bulge ($M_{bul}$) and disc ($M_{disc}$, to allow 
morphological classification of model galaxies. In the following, we 
describe how the evolution of each component is calculated. 

\subsubsection{Gas Cooling \& Reionisation}\label{cool}
\begin{figure} 
\includegraphics[width=0.45\textwidth]{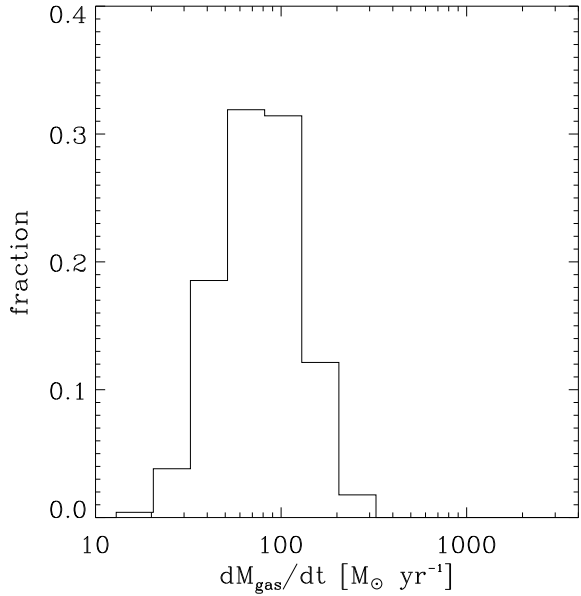}
 \caption{The distribution of SAM accretion rates of cold gas onto central galaxies living in halos with $M_{DM} \sim 10^{12}$ M$_{\odot}$ at $z=2.5$. The distribution is in good agreement with the results from fully numerical simulations presented in \citet{2009Natur.457..451D}. Please note that we do not include major mergers in this plot due to the ambiguity of the time scale over which the gaseous material of the satellite galaxy contributes to the central galaxy. The missing mergers explains the slight mismatch at high accretion rates with  \citet{2009Natur.457..451D}.  } \label{f1}
\end{figure}

Each branch of the merger tree starts at a progenitor mass of $M_{min}$ and 
merges at a redshift $z\geq 0$ with the main progenitor branch. Initially, each halo is occupied by hot 
primordial gas which is captured in the potential well of the halo and shock 
heated to its virial temperature 
$T_{vir}=35.9\left[V_c/(\mbox{km s}^{-1}) \right]^2$ K, where $V_c$ is the 
circular velocity of the halo \citep[K99]{1991ApJ...379...52W}. Subsequently this hot gas component is allowed to radiatively cool and settles down into a
rotationally supported gas disc at the centre of the halo, which we identify 
as the central galaxy \citep[e.g][]{1977ApJ...211..638S,1978MNRAS.183..341W,1991ApJ...379...52W}.  
The rate at which hot gas cools down is estimated by calculating the 
cooling radius inside the halo using the cooling functions provided by 
\citet{1993ApJS...88..253S} and the prescription in S01 assuming the gas density profile follows that of the hosting dark matter at all times. At early times and in low mass haloes the cooling times of the gas are much shorter than the halo dynamical time suggesting that no stable shock will exist at the virial radius of the halo, and that the gas does not shock-heat to virial temperature  \citep{2003MNRAS.345..349B,2005MNRAS.363....2K}. We capture this situation by calculating the radiative cooling time of the halo gas under the assumption of it being at virial temperature and comparing it to the halo dynamical time. Whichever is longer will be used to calculate the cooling rate/accretion rate of gas onto the central galaxy. This procedure has been shown to provide an accurate distinction between haloes with shock heated gas and without around a halo mass scale of $\sim10^{12}$~M$_{\odot}$ \citep{2006MNRAS.365...11C,2008ApJ...680...54K}. Haloes more massive than this show at $ z \gtsim 2$ signs of accretion of cold gas along cosmic filaments \citep{2008MNRAS.390.1326O,2009Natur.457..451D}, which is not captured within the above modelling approach. We here include the approach of \cite{2009ApJ...700L..21K} that models this behaviour and is based on fits to the results from numerical simulations by \cite{2009MNRAS.395..160K}. In Fig. \ref{f1} we show the distribution of accretion rates onto central galaxies in a halo of $M_{DM} \sim 10^{12}$ M$_{\odot}$ at $z=2.5$. The distribution is in good agreement with the results from fully numerical simulations presented in \citet{2009Natur.457..451D}. 

In the case of a merger between halos we assume that all of the hot gas present in the progenitors is shock heated to the virial temperature of the remnant halo, and that gas can only cool down onto the new central galaxy which is the central galaxy 
of the most massive progenitor halo. The central galaxy of the less massive 
halo will become a satellite galaxy orbiting inside the remnant halo. In this 
way, a halo can host multiple satellite galaxies, depending on the 
merging history of the halo, but will always only host one central galaxy onto 
which gas can cool. The cold gas content in  satellite galaxies  is given by 
the amount present when they first became satellite galaxies and does 
not increase, instead it decreases due to ongoing star formation and supernova 
feedback \citep[see however, e.g. ][for models on gradual gas stripping]{2008ApJ...680...54K,2008ApJ...676L.101K}.

In the simplified picture adopted above, the amount of gas available to 
cool down is only limited by the universal baryon fraction. However, in the presence of a photoionising background the fraction of baryons captured in small halos is reduced \citep{2000ApJ...542..535G} and we use the recipe of \citet{2002ApJ...572L..23S}, which is based on a fitting formulae derived from hydrodynamical simulations by \citet{2000ApJ...542..535G}, to estimate the amount of baryons in each halo. For the epoch 
of  reionisation, we assume $z_{reion}=10$, based on recent results from WMAP  \citep{2010arXiv1001.4538K}. 

\subsubsection{Cooling shut-off mass scale}
Cooling rates for massive galaxies must drop off significantly in order to e.g. reproduce the observed colour distribution and mass function of galaxies \citep[e.g.][]{2006MNRAS.365...11C}. Possible candidates for physical processes responsible in shutting down cooling in massive dark matter haloes are e.g. AGN-feedback \citep[e.g][]{2004ApJ...600..580G,2006MNRAS.365...11C,2006MNRAS.370..645B} or gravitational heating by infalling sub-structures \citep{2008ApJ...680...54K,2008MNRAS.383..119D,2009ApJ...697L..38J}. 
Effectively these processes result in cooling being suppressed in haloes above a critical mass scale, and have been implemented using a cooling shut-off in haloes above a critical circular velocity (K99) or mass scale \citep{2006MNRAS.370.1651C}.  By applying an effective shut-off mass scale of $M_{DM,crit}  \approx 10^{12}$ M$_{\odot}$ at $z \ltsim 2$ one is able to reproduce the colour distribution and mass function of galaxies \citep{2006MNRAS.370.1651C}. Additionally, such models are able to reproduce the frequency of dry, gas-poor,  major mergers \citep{2009MNRAS.397..506K}. 
In the following we will modify our fiducial cooling model as laid out in chapter \ref{cool} by adopting a quenching of cooling in dark matter haloes above a critical mass scale of  $M_{DM,crit}\geq 2.5 \times 10^{12}$ M$_{\odot}$ at $z \leq 2$ which is  similar to the value proposed in \citet{2006MNRAS.370.1651C} $(M_{DM,crit}\geq 2 \times 10^{12}$ M$_{\odot})$, and allows to have gas in massive late-type galaxies (see Fig. \ref{f5}).
Note that we allow the gas that is already in the disc of a galaxy to continue forming stars until it is used up, even after the host halo is more massive than $M_{DM,crit}$.

\subsubsection{Star Formation in Discs and Supernova Feedback}\label{sf}
Once cold gas has settled down in a disc, we allow for fragmentation and 
subsequent star formation according to a parameterised global 
Schmidt-Kennicutt type law  \citep{1998ApJ...498..541K} of the form 
$ \dot{M}_{*}=\alpha M_{gas}/t_{dyn,gal}$, where $\alpha$ is a free parameter
describing the efficiency of the conversion of cold disc gas $M_{gas}$ into stars, and $t_{dyn,gal}$ is the dynamical time of the galaxy and is 
approximated to be 1/10 times the dynamical time of the hosting dark matter halo \citep[][K99]{1998MNRAS.295..319M}. For a satellite galaxy we calculate the dynamical time using the halo properties when it last was a central galaxy. We assume here that newly formed stars $\dot{M}_*$ all contribute to the stellar disc $M_{disc}$ of the galaxy.

Feedback from supernovae plays an important role in regulating star 
formation in small mass halos and in preventing too massive satellite 
galaxies from forming \citep{1986ApJ...303...39D,2007ApJ...668L.115K}.
We implement feedback based on the prescription presented in K99 with
\begin{equation}
  \Delta M_{reheat}=\frac{4}{3} \epsilon \frac{\eta_{SN} E_{SN}}{V_{c}^{2}} 
  \Delta M_*. 
\end{equation}
Here we introduce a second free parameter $\epsilon$ which represents our 
lack of knowledge on  the efficiency with which the energy from supernovae 
reheats the cold gas.  The expected number of supernovae 
per solar mass of stars formed for a typical IMF (e.g. Scalo) is  $\eta_{SN}=5 \times 10^{-3}$,  and $E_{SN}=10^{51}$ erg  is the energy output from each supernova. We take 
$V_c$ as the circular velocity of the halo in which the galaxy was  
last present as a central galaxy or is right now.

\subsubsection{Implementation}
In between mergers the evolution of galaxies is governed by the equations presented in the previous sections. We solve this coupled system of differential equations by integration in between time steps of constant $\Delta z = 0.02$, during which the halo properties are kept fixed. We thus solve following set of equations:
\beq
\dot{M}_{*} & = & \alpha \frac{M_{gas}}{t_{dyn,gal}} \nonumber \\ 
\dot{M}_{reheat} & = & \frac{4}{3} \epsilon \frac{\eta_{SN} E_{SN}}{V_{c}^{2}}  \dot{M}_*   \\
\dot{M}_{gas} & = & \dot{M}_{cool} - \dot{M}_{reheat} - \dot{M}_*  \nonumber\\
\dot{M}_{hot} & = & - \dot{M}_{cool} + \dot{M}_{reheat} \nonumber \\
\dot{M}_{disc} & = & \dot{M}_{*}  \nonumber.
\eeq

Here $\dot{M}_{cool} $ is the cooling rate applying the prescriptions of the previous sections. The above set of equations is strictly only applicable to central galaxies, which show cooling of hot gas. For satellite galaxies we use a modified set of equations with $ \dot{M}_{hot} = M_{hot} = \dot{M}_{cool} = 0$. All gas that is reheated in satellites is added to the hot gas reservoir of the central galaxy.

\subsubsection{Galaxy Mergers and Bulge Formation}\label{merger}
We allow for mergers between galaxies residing in a single halo. As mentioned 
earlier, each halo is occupied by one central galaxy and a number of 
satellite galaxies depending on the past merging history of the halo. 
Whenever two halos merge, the galaxies inside  them will merge on a 
time-scale which we calculate by estimating the time it would take the 
satellite to reach the centre of the halo under the effect of 
dynamical friction. Satellites are assumed to merge only with 
central galaxies and we set up their orbits in the halo according to the 
prescription of K99, modified to use the Coulomb logarithm 
approximation of S01 and the correction factor of \citet{2008ApJ...675.1095J}. 
Expressed in units of the Hubble time the dynamical friction is only dependent on the mass ratio of the merging partners. The merger rates in our SAM are in good agreement with observations \citep{2009ApJ...697.1971J} and agree within a factor of few with other models \citep{2010ApJ...715..202H,2010arXiv1004.2708H}.
 
We do not distinguish between the effects of major $R \equiv M_{gal,1}/M_{gal,2} 
\leq 3.5$ ($M_{gal,1} \geq M_{gal,2}$) and minor mergers $10 > R > 3.5$ in our SAM, but rather apply a model that shows a smooth transition between the effects of such mergers. In what follows the index 1 will always refer to the primary galaxy and the index 2 to the merging satellite galaxy.  During major mergers the stellar discs of the progenitors will be almost completely destroyed and added to the bulge component of the remnant. In practice we assume that the amount of disc mass  scattered into the bulge is min$[M_{disc,1},M_{*,2}]$  \citep{2009ApJ...691.1168H}.  In addition we assume that a star burst will be triggered due to the interaction of the galaxies. The fraction of cold disc gas transformed into stars during the star burst is modelled using a function $f_{burst} $ that includes dependencies on the cold gas fraction $f_{gas} \equiv M_{gas}/(M_{gas}+M_*)$, the disc fraction $f_{disc} \equiv M_{disc} / M_*$ and mass ratio  $R$ of the merging galaxies \citep{2009ApJ...691.1168H}. We here use the following functional form that shows broad agreement with results from detailed numerical merger simulations \citep{2009ApJ...691.1168H}.
\beq
f_{burst} = 1 - (1+r_{sb} / r_{disc}) \exp(-r_{sb}/r_{disc}).
\eeq
We calculate the disc scale length $r_{disc}$ using the disc model of \citet{1998MNRAS.295..319M} and the radius $r_{sb}$, within which gas loses its angular momentum and contributes to the central star burst via:
\beq
r_{sb} = \alpha_{sb} (1-f_{gas}) f_{disc} F_\theta G(R),
\eeq
with $G(R) \equiv 2R/ (1+R)$. We set the numerical constant $\alpha_{sb}=1$ and the orbit dependent constant $F_\theta=1.2$ \citep{2009ApJ...691.1168H}. The latter represents an average value for a range of different merging orbits.  
 
The cold gas that is not converted into stars during the merger is added to the gaseous disc of the remnant. This results in following composition for the remnants of major and minor mergers:
\beq\label{eq2}
M_{bul,rem} &=& M_{bul,1} + M_{bul,2} + \min[M_{disc,1},M_{*,2}]  \nonumber \\ 
 			  & & + M_{disc,2} + f_{burst,1} M_{gas,1} \nonumber \\
			  & &	 +  f_{burst,2} M_{gas,2}   \\
M_{disc,rem} &=& M_{disc,1}-\min[M_{disc,1},M_{*,2}]   \\
M_{gas,rem} &=&  (1-f_{burst,1}) (M_{gas,1})   \nonumber \\
			& & + (1-f_{burst,2}) M_{gas,2}). 
\eeq

Immediately after major mergers our model galaxies have a large fraction of their stars in the  bulge component and the stellar disc can start re-growing by either conversion of cold gas that was not used up in the merger or by freshly accreted gas. We here assume mergers as the only efficient means of bulge formation and neglect other possible secular effects such as e.g. disc instabilities  \citep[see e.g.][]{2009ASPC..419...31C}.

\section{Model normalization}\label{modnor}
The various prescriptions for physical process introduced in the last section include a number of free parameters which represent our lack of knowledge on parts of the underlying physical processes. Generally SAMs use a set of observations to fix these parameters. Reproduction of such observations thus cannot been seen as a model prediction but rather a  'tuning' of it. In this section we present the core observations our free parameters are tuned to and their individual values. 
\subsection{Mass function}
\begin{figure} 
\includegraphics[width=0.45\textwidth]{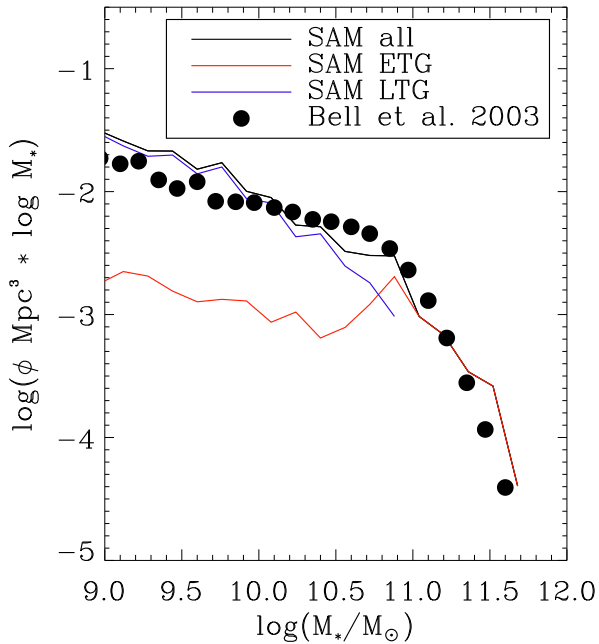}
 \caption{The mass function of galaxies at $z=0$ in our best-fit model (solid line). Filled circles are measurements by \citet{2003ApJS..149..289B}. Red and blue lines show the early-type and late-type galaxy mass function, respectively. }\label{f3}
\end{figure}

The mass function of galaxies is a core observable that is well measured and robust. Much progress in galaxy modelling has been achieved over the last years by focusing on reproducing specific features in it such as e.g. the steep exponential decline at high masses \citep{2003ApJ...599...38B,2006MNRAS.365...11C,2008ApJ...680...54K} or the steepness of the faint-end slope as a function of redshift \citep{2007ApJ...668L.115K}. The \atl parent sample (Paper I) shows very good agreement with the mass function of \citet{2003ApJS..149..289B}, to which we compare our best-fit $z=0$ model mass function (Fig. \ref{f3}). We find very good agreement both in shape and overall normalisation over a wide range in masses, in particular at the high mass end, where our model population is dominated by early-type galaxies with bulge-to-total mass ratios $B/T \gtsim 0.7$ (see Fig. \ref{f4}).

\subsection{Bulge-to-Total Mass Ratios}
\begin{figure} 
\includegraphics[width=0.45\textwidth]{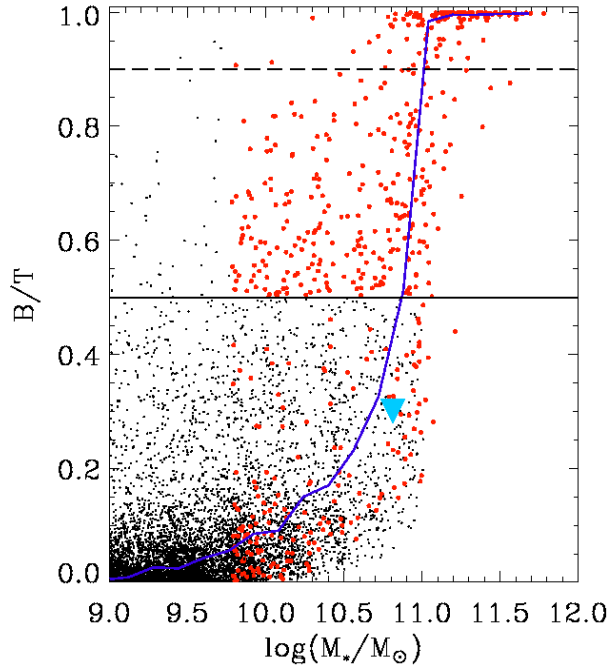}
 \caption{Bulge-to-total mass ratios $B/T$ of galaxies as a function of mass. The solid blue line shows the median value at a given galaxy mass. Our results show a clear trend towards bulge-dominated galaxies at larger masses. The overall scatter of the distribution is largest around $10^{11}$ M$_{\odot}$, which also marks the point at which the galaxy populations is largely dominated by early-type galaxies. The solid triangle shows the Milky-Way value of \citet{2007MNRAS.379..755S}. The horizontal solid and dashed line mark the transition to early-type galaxies $(f_{early}=0.5)$ and slow rotators $(f_{fast}=0.1)$, respectively. Large red circles show the final  extended early-type galaxy sample in the SAM, including the additional criterion $f_{red}= M_{gas}/M_* =0.15$, on the minimum gas fraction in late-type galaxies. The latter has been included to mimic the \atl selection of early-type galaxies, which allows for the inclusion of gas-poor low $B/T$ galaxies.  }\label{f4}
\end{figure}

Continued merging is transforming morphologies of galaxies by scattering disc stars into bulges. If modelled correctly, the observed trend of increasing bulge fractions with galaxy mass should be recovered by the model. Since we focus here on early-type galaxies this is an important test. In Fig. \ref{f4} we show the $B/T$ and stellar masses of our model galaxies, as well as the median $B/T$ at a given mass.  We are able to recover the general observed trend toward bulge dominated systems with increasing mass. The median $B/T$ is below 0.2 for galaxies with $M_* \ltsim 5 \times 10^{10}$ M$_{\odot}$ and makes a strong transition to $B/T > 0.6$ at $M_* \gtsim 10^{11}$ M$_{\odot}$. It appears when the cooling of gas starts slowing down, in haloes below $M_{DM,crit}$. As a consequence discs grow slower, while merging continues to transform discs into spheroids, shifting the median $B/T$ to larger values. 
\subsection{Gas Fraction}
\begin{figure} 
\includegraphics[width=0.45\textwidth]{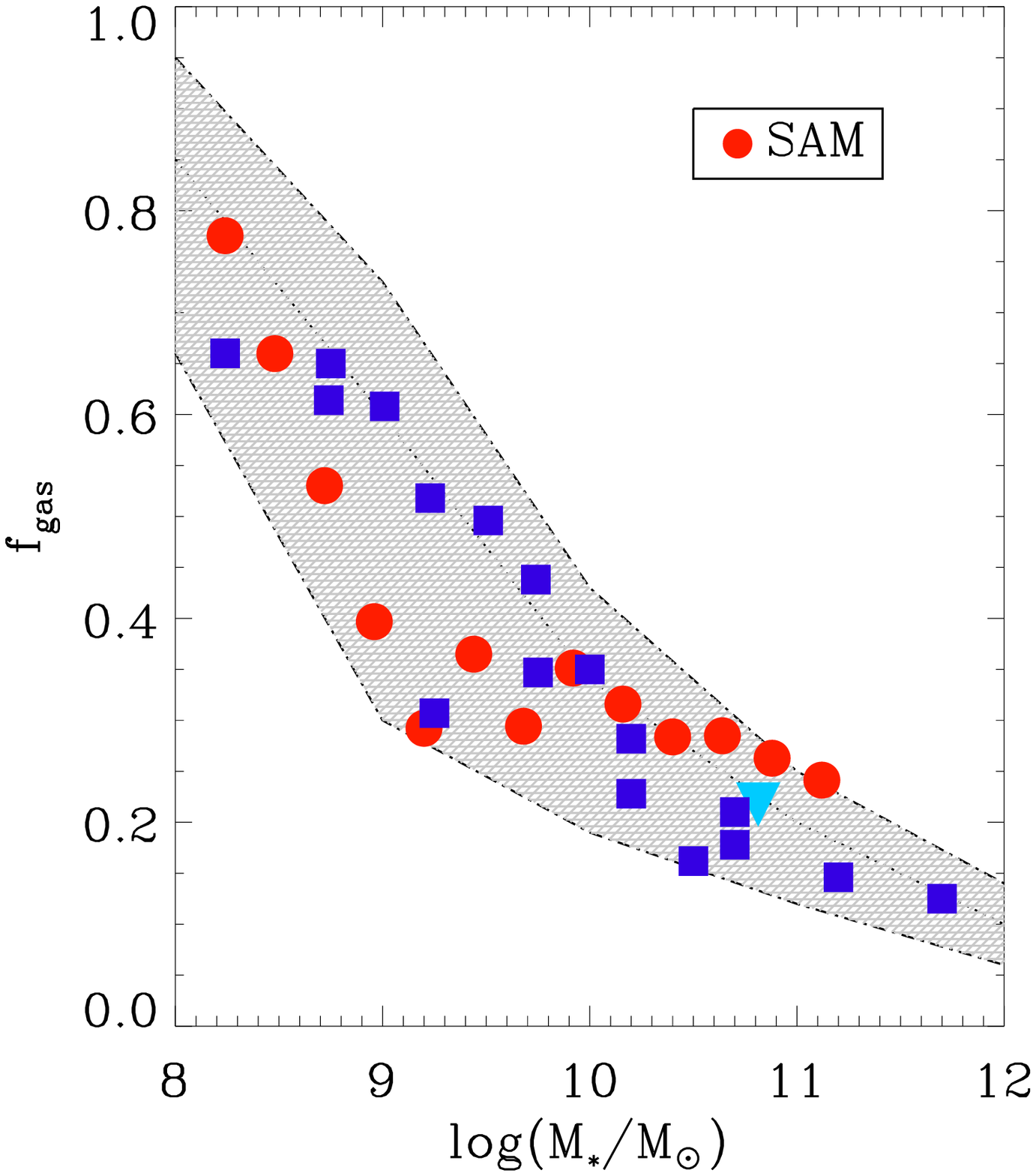}
 \caption{The median fraction of gas $f_{gas}= M_{gas}/(M_{gas}+M_*)$ as a function of galaxy mass in the SAM (filled circles) range covered by observations including their error bars (hatched region). Solid squares show the compilation of various observational data \citep{2001ApJ...550..212B,2004ApJ...611L..89K,2005ApJ...632..859M}, and the solid triangle is the value for the  Milky-Way. The declining gas fraction with stellar mass is mainly due to the increasing cooling time scales for the gas. Please note, that due to the sharp cooling shut-off in halos above $M_{DM,crit}$ the gas fraction declines very sharply  to zero.  }\label{f5}
\end{figure}

Another important quantity related to the efficiency of disc re-growth and survival during mergers is the cold gas fraction, $f_{gas} \equiv M_{gas}/(M_{gas}+M_*)$,  as a function of the stellar mass. We show the median value from our model for late-type galaxies with $B/T < 0.5$ in Fig. \ref{f5}, and compare it to the observed range in disc galaxies \citep{2001ApJ...550..212B,2004ApJ...611L..89K,2005ApJ...632..859M}. We find fair agreement for model galaxies with $M_* > 10^9$ M$_{\odot}$ and the observations. The decreasing gas fraction in our model is mainly due to the increasing cooling time as a function of host halo mass, and the sharp cooling shut-off in haloes above $M_{DM,crit}$.

\subsection{Early-Type classification}\label{elld}
In general SAMs separate between early-type and late-type galaxies using the bulge-to-total ratio $B/T$ either in mass or $B-$band magnitude \citep[e.g.][]{1999MNRAS.303..188K,spr01,2006MNRAS.370..645B}. This approach is based on the good correlation between the Hubble T-type and the  $B-$band  $B/T$ ratio of galaxies \citep{1986ApJ...302..564S}. The early-type galaxies selected within the \atl survey are in very good agreement with morphologies based on the T-type classification scheme (Paper I). 
We will use here a fiducial value for the bulge-to-total mass ratio of  $B/T =0.5$ to separate early-type from late-type galaxies. 

The \atl survey however, contains a significant numbers of flat disc-like galaxies, that have morphologies similar to late-type spirals, but do not show evidence for gas, dust or spiral arms, and are for this reason classified as early-types \citep{pap7} according to the standard morphological classification \citep{Sandage1961}. Such {\it red} spirals are classified as fast rotators \citep[][hereafter Paper II and Paper III, respectively]{pap2,pap3}  and are not included, if the early-type selection in the SAM is done purely based on a $B/T$ cut. The combination of a strong decline in the mass function of galaxies and the transition from fast to slow rotator dominance, suggest that the impact on the ratio of fast to slow rotators by missing such galaxies, could be  important at masses $\gtsim 10^{11}$ M$_{\odot}$ (Paper III). We address this issue by including an additional criterion based on the gas fraction in galaxies with low $B/T$. We first compare the distribution of the cold HI gas fraction $f_{red} \equiv M_{gas}/M_{*}$ in model central galaxies with that obtained from the \atl sample \citep{paolo} assuming a constant molecular-to-atomic ratio \citep[see however,][for more detailed modelling of the molecular fraction in cold gas]{2009ApJ...698.1467O}. 
The distributions show very similar trends with a peak at high gas fractions for late-type galaxies and ETGs spanning a wide range from low to high  gas fractions with a strong tendency for low gas fractions \citep{paolo}. Modelled late-type galaxies ($B/T < 0.5$) show in addition a small peak at gas fractions below  $\sim 15 \%$ which is not seen in the observations
 Such galaxies are likely candidates for fast rotators with low $B/T$ and we therefore adopt an additional criterion for early-type galaxies with $B/T <0.5$ and $ f_{red} < 0.15$.  A significant fraction of galaxies with masses   $ \sim 10^{11} $M$_{\odot}$ in our model are centrals living in haloes which have been crossing $M_{DM,crit}$ some time ago, and since then did not encounter any significant mergers thus using up their gas reservoir and building a stellar disc with little interference.  The hosting halo of these galaxies are not very much larger than   $M_{DM,crit}$, thus explaining why they do not reach masses larger than $\sim 10^{11}$ M$_{\odot}$. In general we find that the majority of low $B/T$ fast rotators below $ \sim 10^{11} $M$_{\odot}$ are satellite galaxies. 
The fraction of early-type galaxies is known to increase with the density of the environment \citep{1980ApJ...236..351D}. A consistent trend has been found for the galaxies in the \atl survey \citep[hereafter Paper VII]{pap7}. To include to first order gas-poor low $B/T$ satellite galaxies in high density environments, we add satellite galaxies in group-like dark matter haloes with $M_{DM} \gtsim 4 \times 10^{13}$  M$_{\odot}$ and stellar masses larger than $10^{10}$ M$_{\odot}$. Again we here use the early-type definition with $B/T < 0.5$ and $f_{red} < 0.15$ for these satellite galaxies. We find that the trend in the ratio of fast to slow rotators is not significantly affected by adding the satellite population, as most of them have masses in a range where fast rotators clearly dominate the population of early-type galaxies. 

In Fig. \ref{f4} we show the final selection of early-type galaxies from the overall model sample. This selection yields a total number density of $ 1.6 \times 10^{-3}  $Mpc$^{-3}$ early-type galaxies with $M_* \geq 6 \times 10^9$ M$_{\odot}$, which is in good agreement with the observed value for the \atl sample of $2.2 \times 10^{-3} $ Mpc$^{-3}$ (Paper I).
In Fig. \ref{f3} we show the mass function separated into early-type and late-type galaxies for our adopted selection. The high-mass end is as expected dominated  by ETGs, and we find that the number density is nearly constant for $M_* < 10^{11}$ M$_{\odot}$. Compared to the observed mass function presented in Paper I our model produces too few ETGs at low masses, most of which should be fast rotators and too few very massive $M_* > 10^{11}$ M$_{\odot}$ late-type galaxies.

 Approximately $80 \%$ of the model early-type galaxies have $B/T > 0.5$ while the remaining $ 20 \%$ are low $B/T$ galaxies devoid of gas.

\begin{table}
\centering
\begin{tabular}{l|l|l|l}
\hline
\hline
star formation efficiency & $\alpha$ & 0.02 & $\S$ \ref{sf} \\
supernovae feedback efficiency & $\epsilon$ & 0.4 & $\S$ \ref{sf} \\
minimum gas fraction in late-types & $ f_{red} $ & 0.15 & $\S$ \ref{elld} \\
early-type threshold & $ f_{early} $ & 0.5 & $\S$ \ref{elld} \\
fast rotator threshold & $ f_{fast} $ & 0.1 & $\S$ \ref{fast} \\
\hline
\hline
\end{tabular}
\caption{Most important free model parameters and their adopted values.}
\label{tab}
\end{table}

\section{ Fast and Slow rotating elliptical galaxies}\label{fas}
The results of the \atl survey suggest that ETGs can be classified into two classes based on the parameter  $\lambda_R \equiv \langle  R \left|V \right| \rangle / \langle R \sqrt{V^2+ \sigma^2} \rangle $ which has been shown to be a good proxy  for the specific baryonic angular momentum  in ETG \citep[Paper III]{2007MNRAS.379..401E,2009MNRAS.397.1202J}. Using a suite of high resolution simulations Paper VI shows that during major mergers ($R < 2$) of late-type galaxies the geometry of the encounter is the dominant factor in deciding whether the remnant is a slow or fast rotator. In general, major mergers that are retrograde (prograde) with respect to the main progenitor result in slow (fast) rotators. In minor mergers ($R \ge 3$) remnants are almost exclusively fast rotators. However, the situation is not as simple as that. Simulated slow rotators tend in general to be too flat compared to observed ones and have properties similar to $2\sigma$ galaxies (Paper II), suggesting that a single major merger might not be the way to form them. Furthermore, merging of remnants show that the orbital angular momentum is able to spin up slow rotators, even in major mergers. Thus  to predict the proper outcome of a binary merger, one needs to know the initial state of the progenitors in terms of their angular momentum and how the orbital angular momentum is re-distributed within the remnant.  Ideally one would follow individual galaxies and calculate the change in $\lambda_R$ due to mergers and star formation in discs, the latter generally contributing to an increase in $\lambda_R$. 

To assess the impact of mergers on the evolution of $\lambda_R$ one can to first order use the analogy to dark matter haloes and their spin parameter $\lambda$ \citep{2007MNRAS.379..401E}. Following the merging history of dark matter haloes, and under the assumption of random orbits and complete transfer of orbital angular momentum to the remnant, one recovers a log-normal distribution for $\lambda$ consistent with results from cosmological simulations \citep{2002MNRAS.329..423M}. The distribution of $\lambda$ in dark matter haloes shows only a weak mass dependence \citep{2007MNRAS.376..215B}, in stark contrast to the strong mass dependence of $\lambda_R$ in ETGs implying that whatever drives the trend in $\lambda_R$ with mass is not just simple transfer of orbital angular momentum. While individual mergers might burst or lower $\lambda_R$ occasionally based on the orbit (Paper VI), continuous merging on random orbits will drive $\lambda_R$ to a mass independent distribution at a median value of $\lambda_R \sim 0.038$ \citep{2002MNRAS.329..423M}. Here we used the relation between the spin parameter $\lambda$ and $\lambda_R$ presented in \citep{2007MNRAS.379..401E} and the median spin parameter of dark haloes in \citet{2007MNRAS.376..215B}. Such low $\lambda_R$ values would classify the majority of ETGs as slow rotators, thus additional physical processes need to be important.

Detailed kinematic analyses of \atl galaxies by Paper II shows, that a large fraction of fast rotators have disc-like regular rotation patterns and bars, suggesting that the stellar body is hosting a disc component. Stellar discs in terms of  $\lambda_R$  would all be classified as a fast rotating component (Paper VI). If mergers indeed on average lead to low $\lambda_R$, the presence of a disc can be all that makes the difference between fast and slow rotators. The process of disc building (increasing $\lambda_R$) is in direct competition with the process of disc destruction by mergers (lowering $\lambda_R$).   

Motivated by above arguments we introduce a simple model based on the mass fraction in discs to investigate the origin of fast and slow rotating ETGs. In what follows we choose the fraction of fast-rotators as a function of luminosity as our key-observable, and gauge our models against it. 

\subsection{Fast Rotators} \label{fast}
\begin{figure} 
\includegraphics[width=0.45\textwidth]{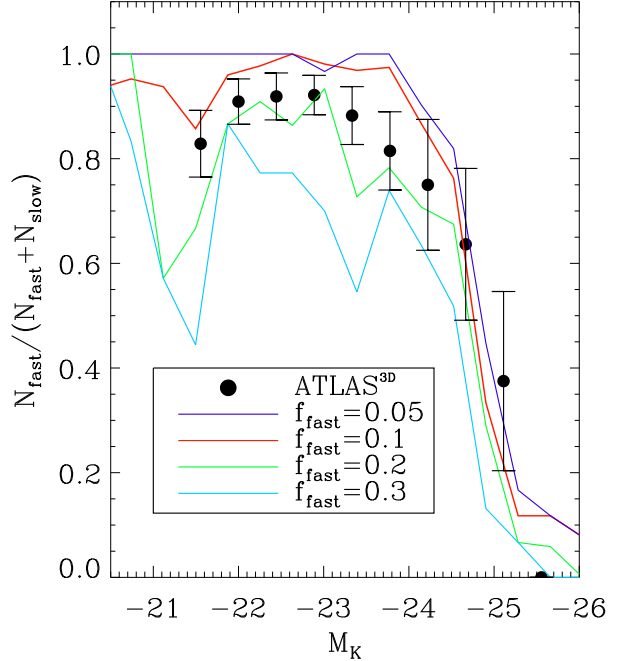}
 \caption{The fraction of fast rotators in the overall population of early-type galaxies in the \atl sample (filled symbols) (Paper III). The SAM outputs for different values of $f_{fast}$ are shown as  different coloured solid lines. Throughout this paper we will use as our fiducial value for the minimum disc fraction in fast rotators $f_{fast}=0.1$.   } \label{f6}
\end{figure}

Fast rotators are the majority of early-type galaxies and dominate the population at $M_K \gtsim -24$ (see Fig. \ref{f6}). Detailed analyses of several fast rotators by \citet{2008MNRAS.390...93K,pap2} reveal the existence of disc-like features in early-type host galaxies in agreement with earlier studies of  \citet{1990ApJ...362...52R}. We here assume in agreement with Paper VII that such stellar discs are present in all fast rotators and use their existence as our criterion to distinguish between fast and slow rotators. We introduce a free model parameter $f_{fast}$, that gives the lower limit to the disc mass in fast rotators. Every early-type model galaxy with a disc fraction $M_{disc}/M_* < f_{fast}$ is classified as a slow rotator.  
We calculate $K-$band magnitudes for our model galaxies using their star formation history and the stellar population synthesis models of \citet{2003MNRAS.344.1000B}.
In Fig. \ref{f6} we show the fraction of fast rotators in the overall population of early-type galaxies as a function of their rest-frame $K$-band magnitude for the \atl sample. The sample shows a clear trend toward slow rotators at the luminous end and is dominated over the remaining observed range by fast rotators (Paper III). 
 We show the fraction of fast rotators from our SAM for different values of the free parameter $f_{fast}$. In general we find that values of $f_{fast} \sim 0.1 - 0.15$ show good agreement with the observations. We here choose $f_{fast}=0.1$ as our fiducial value, based on matching the observational data in Fig. \ref{f6} and the additional constraint of being close the number density of fast and slow rotators from the \atl sample (see Fig. \ref{f10}). 
 The exact value of $f_{fast}$ does not influence the shape of the curve shown in Fig. \ref{f6} it only changes the magnitude at which fast rotators start to dominate. The modelled galaxy population of fast and slow rotators is in good agreements with the observations, supporting the idea that stellar discs are the dividing factor between fast and slow rotators.   

The formation path toward fast rotating ETGs in our model is either by rebuilding a stellar disc around a spheroid, the growth of a spheroid via the destruction of existing discs in  late-type galaxies or the exhaustion of gas in late-type galaxies.  Slow rotators on the other hand have two possible formation paths: major mergers or repeated dry minor mergers. In what follows we will investigate the importance of these different paths by changing the model prescription within our fiducial SAM. We will focus in particular on the importance of gas cooling, the destruction of stellar discs in major mergers, star burst during major mergers and the effects of minor mergers. 

\subsubsection{Impact of Gas Cooling}\label{gc}
We first change our fiducial model in a way that highlights the importance of gas cooling for the formation of fast rotators.
We will call it  {\it max-merg} in the following (label {\it MM} in Fig. \ref{f7} \& Fig. \ref{f7b}). In this model every binary major merger with $R \leq 3.5$  results in a slow rotator, and all the progenitor disc mass is added to the bulge, as well as all progenitor gas is converted into bulge stars during a star burst. The modification to the fiducial model (Eq. \ref{eq2}) is summarised below. For major mergers we assume:
\beq \label{eq3}
M_{bul,rem} &=& M_{bul,1} + M_{bul,2} + M_{dis,1} + M_{dis,2} \nonumber \\ 
 			  & & +  M_{gas,1} + M_{gas,2}  \nonumber \\
M_{dis,rem} &=& 0   \\
M_{gas,rem} &=& 0 \nonumber 
\eeq 
and for minor mergers with $10 \geq R > 3.5$:
\beq \label{eq4}
M_{bul,rem} &=& M_{bul,1} + M_{*,2}  \nonumber \\ 
M_{dis,rem} &=& M_{dis,1}     \\
M_{gas,rem} &=&  (M_{gas,1} + M_{gas,2}). \nonumber 
\eeq
In this model the formation of fast rotators via discs is  maximally hindered and only allowed via cooling of gas from the hot halo and cold gas from gas-rich satellites in minor mergers. The solid  blue line in Fig. \ref{f7} shows the fraction of fast rotators that formed via disc growth around bulges. We find the expected trend of this mode being more important at lower mass scales, when cooling times are shorter, most visible at  $M_K \gtsim -24$. However, shorter cooling times also mean that more gas will have cooled down before a merger takes place, and thus less material is available to re-grow a disc. For massive galaxies one can use a rough estimate to evaluate the role of cooling with respect to mergers. Observations of the merger rate in the nearby universe find on average $0.0001 $Gyr$^{-1}$ Mpc$^{-3}$ mergers for massive galaxies with $M_* > 2.5 \times 10^{10}$ M$_{\odot}$ \citep{2009ApJ...697.1971J}. For the \atl early-type sample this suggests less than one merger within the last Gyr per galaxy. Cooling time scales on the other hand are around 1 Gyr \citep{2010MNRAS.405.2717N},  suggesting that cooling processes and associated star formation in disc-like components are not negligible in between merger events. However, cooling by itself is not sufficient  to produce enough fast rotators by means of transforming bulge dominated major merger remnants as evidenced by the large drop in the fast rotator fraction at $M_K \gtsim -24$. The more important contribution to the fraction of fast rotators comes from gas-poor low $B/T$ satellite galaxies living in dense environments (see Fig. \ref{f7b}). On average, fast rotators that are 'dried-up' late-type galaxies in the range $ -22 > M_K > -25$  are $\sim 3$ times more frequent than 're-grown disc'  fast rotators showing that disc re-growth  after a major merger is a less important channel for the formation of fast rotators than possible environmental effects  leading to the exhaustion of gas in late-type galaxies.

\subsubsection{Gas-rich Mergers} 
Within our fiducial merger model stellar discs have two distinct origins. They are either the remains of a stellar disc that survived a merger or are re-grown from gas that settled into a disc after a merger. The gas for the latter process can be part of the progenitor discs that survived the merger or gas that is radiatively cooling and settling onto a disc. The importance of the latter process has been addressed in the last section and we will turn now to the survival of gas in progenitor discs. Numerical simulations show  that during mergers of galaxies with high gas fractions, progenitor discs are able to survive and to start quickly rebuilding stellar discs \citep{2006ApJ...645..986R,2009ApJ...691.1168H}. Our fiducial merger model in Eq. \ref{eq2} does take the results from such simulations into account via the function $f_{burst}$. In the {\it max-merg}  model above we assumed $f_{burst}=1$. To estimate the importance that the survival of gaseous disc during major mergers has on rebuilding stellar discs we modify the max-merg model by using $f_{burst}$ from Eq. \ref{eq2}. The survival of gaseous disc material during  mergers is mainly affecting the fraction of fast rotators between $M_K \sim -23$ and $M_K \sim -24$, with only a mild increase in the total number of fast rotators by a factor of $1.6$ (see Fig. \ref{f7} \& Fig \ref{f7b} dotted line, label {\it MM\_G}). From this we conclude,  that in our model the formation  of fast rotators via the growth of stellar discs from cold disc gas that is surviving major mergers is less important than via radiative cooling of gas. 
 
\subsubsection{Survival of Stellar Discs}
\begin{figure} 
\includegraphics[width=0.45\textwidth]{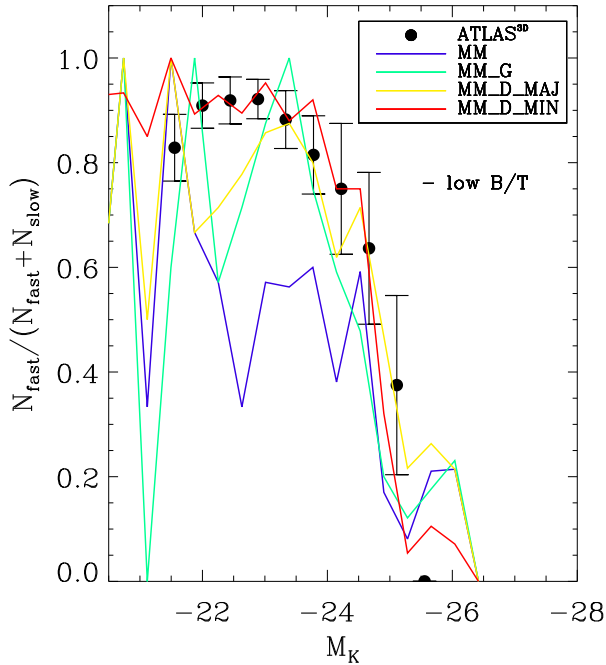}
 \caption{The fraction of fast rotators in the overall population of early-type galaxies in the \atl sample (filled symbols) (Paper III). The SAM outputs for different model assumptions are shown by the various lines in the figure (see text for details on the models). For all models we use $f_{fast}=0.1$. The blue, green and yellow lines show models including the effects of major mergers only, while the red line shows the outcome for a model including minor mergers. Results in this figure are shown for fast rotators that do not have $B/T < 0.5$ ({\it - low B/T}). The survival of stellar and gaseous discs during major mergers (models MM\_D\_MAJ and MM\_G, respectively) plays mostly a role for fast rotators around $M_K \sim -23$, while the overall population of fast rotators is dominated by regrowth of a disc from gas cooling and disc destruction in minor mergers.  } \label{f7}
\end{figure}

Detailed numerical simulations of binary galaxy mergers with and without gas show, that the mass ratio of the merging partners plays a crucial role in erasing the memory of the progenitor's stellar disc  \citep[e.g.][]{1992ARA&A..30..705B,2003ApJ...597..893N,2009ApJ...691.1168H}. Within our fiducial model we always destroy a fraction of the stellar disc that is given by $\min[M_{di,1},M_{*,2}]$. Effectively, our model states that ten $10:1$ mergers have the same effect on a stellar disc as  one $1:1$ merger \citep{2007A&A...476.1179B}. To see how the survival of a stellar disc will influence the fraction of fast rotators we modify the max-merg model to include the dependence on the mass ratio  as laid out in our fiducial merger model Eq. \ref{eq2}. Please note that we assume that only in major mergers, $R < 3.5$, the disc of the host galaxy will be affected. Minor mergers are neglected in terms of their impact onto the stellar disc, and we will come back to it in the next section. As can be seen from the dashed line labelled {\it MM\_D\_MAJ} in Fig. \ref{f7}, the fraction of fast rotators is mainly increased for luminous galaxies around $ M_K \ltsim -23$. This increase is modest with respect to the max-merg model and of the same magnitude as in the case  of  including stellar disc survival during mergers. Unequal mass major mergers dominate the overall number of mergers \citep[e.g.][]{2001ApJ...561..517K}, thus the survival of stellar discs is elevated with respect to the max-merg model. Again we find that disc re-growth via cooling is more important for the formation of fast rotators after a major merger than the survival of pre-existing properties during this major merger, in this case the stellar disc of the main progenitor. 

 Our results here and in the previous section show that surviving discs, stellar and gaseous,  are important in allowing to maintain and re-grow substantial discs from cooling of gas in ETGs mostly around $M_K \sim -23$, and that the majority of fast rotators is likely to originate from a different formation path within our model which. 

\subsubsection{Minor Mergers}\label{minor}
\begin{figure} 
\includegraphics[width=0.45\textwidth]{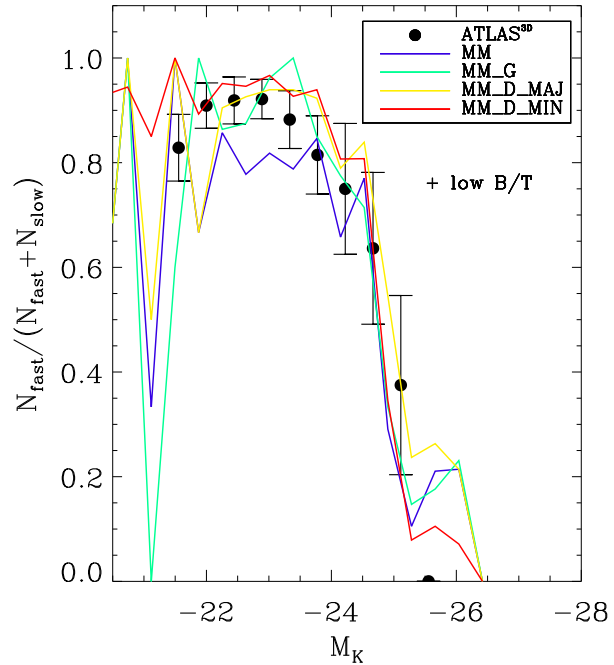}
 \caption{Same as Fig. \ref{f7} but now including the contribution from gas-poor low $B/T$ galaxies ({\it + low B/T}). Such galaxies play an important role in major merger only models. However, in the model including the effect of minor mergers (MM\_D\_MIN) the fraction of fast rotators originated from drying out of late-type galaxies in high density environments is significantly lowered. } \label{f7b}
\end{figure}

The frequency of minor mergers an ETG encounters during its evolution by far exceeds the number of major mergers \citep[e.g.][]{ks06,2009MNRAS.394.1713K,2009ApJ...699L.178N,2010MNRAS.405..948S}. The importance of minor mergers for the formation of bulges in late-type galaxies has been recently pointed out \citep{2009ApJ...696..411W,2009MNRAS.397..802H} and we here investigate their importance in terms of fast and slow rotating ETGs. 
As discussed above, minor mergers are able to disrupt parts of the stellar disc of their host and to trigger a small star burst \citep{2010MNRAS.405.2327P}. We here modify the max-merg model to include the disruption of stellar discs in minor mergers $10 > R > 3.5$, but do not allow for any star burst during it. The solid red line labelled {\it MM\_D\_MIN} in Fig. \ref{f7} shows the outcome of this model. It is clearly visible how the fraction of fast rotators is significantly increased at $M_K \gtsim -25$, with respect to the max-merg model. 
Our results suggest that minor mergers are efficient means in gradually transforming disc dominated systems into systems with $0.5 < B/T < 0.9$, corresponding to the $B/T$ range of fast rotating ETGs. As we will show below, the continued impact of such minor mergers is leading to even higher $B/T$ and the formation of slow rotating ETGs.
 By comparing Fig. \ref{f7} and \ref{f7b} we find that the fraction of fast rotators with such an origin is much larger than the one for fast rotators that are dried up late-type galaxies.  

\subsection{Slow rotators}
\begin{figure} 
\includegraphics[width=0.45\textwidth]{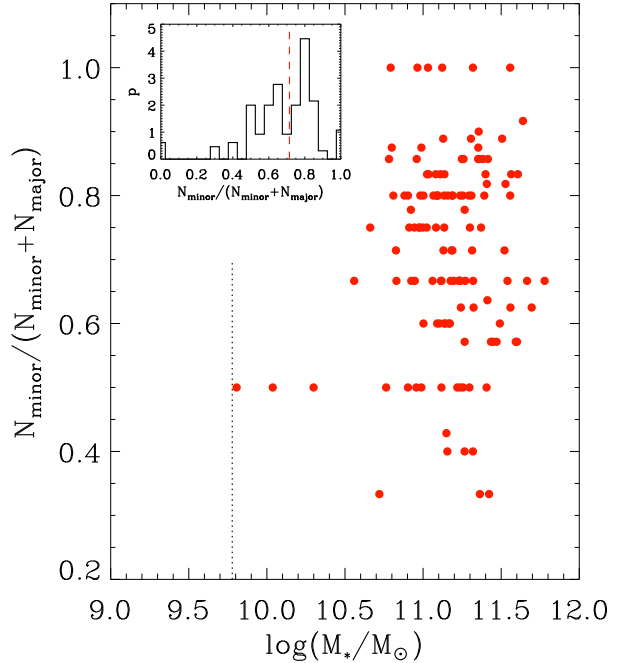}
 \caption{The fraction of minor mergers $N_{minor}/(N_{minor}+N_{major})$ that the most massive progenitor branch of a present-day slow rotator experienced during its evolution. For the sake of better statistics we here define major merger as $R \le 4$ and minor mergers as $20 \ge R >4 $.  The histogram shows the distribution of the minor merger fraction which has a median value of $0.73$ (dashed line). Minor mergers clearly dominate the history of slow rotators. However, we do not find a strong correlation with stellar mass. Slow rotators with very few or no major mergers have more minor mergers than  slow rotators with some major mergers. The evolution paths of these two classes of slow rotators is very different with the former going through a fast rotator phase, while the latter always stays close to the slow rotator regime. The vertical dotted line shows the completeness limit of the \atl sample (Paper I) } \label{f9}
\end{figure}

Mergers play by construction an important role in the formation of ETGs in our model. Our assumption of equal mass mergers predominantly resulting in slow rotators is clearly an oversimplification of what is happening in reality (Paper VI). In general, the formation of a slow rotator requires that the orbital angular momentum and the spin of the galaxies compensate each other such that only little  angular momentum is left in the remnant, a situation that requires fine-tuning, given that the ratio of orbital angular momentum to spin is on average larger than one \citep{2006A&A...445..403K,2002MNRAS.329..423M} and that  slow rotators after one additional major merger tend to spin up to become fast rotators again (Paper VI). 
The question thus arises how slow rotators might form in the first place, if even under favourable conditions, such as equal mass mergers, their formation/survival chances are small \citep{2010MNRAS.tmp..915B}. 
As we pointed out earlier, minor mergers are very frequent and can contribute a significant fraction of the remnant mass. In an earlier study \citet{ks06} showed that up to $80 \%$ of the bulge mass can come from outside the galaxy via mergers, estimates in agreement with numerical simulations of \citet{2010arXiv1010.1381O}. 
If this is indeed the case then slow rotators are the end-stage of constant impact by satellite mergers, falling in from random directions and helping to reduce the angular momentum of the host on average. 
Since the cooling of gas has mostly ceased in high mass galaxies living in haloes $> M_{DM,crit}$, disc formation will be terminated as well, and thus the conversion into fast rotators. 
To highlight the relative importance of minor and major mergers we show in Fig. \ref{f9} the fraction of minor mergers that the most massive progenitors of a present-day slow rotators experienced during their evolution. We here define major and minor  mergers as $R \le 4$ and $20 \le R > 4$, respectively, to increase the statistical significance. 
There is a large scatter in the fraction of minor mergers that slow rotators experienced during their evolution and no significant correlation with galaxy mass. 
The median fraction of minor mergers that the most massive progenitor of a slow rotators experiences is $\sim 0.7$ corresponding to $\sim 2.3$ more minor than major mergers. The implication of these results is that massive slow rotators had most likely a major merger event in their past, but got constantly hit by minor mergers, that erased any progenitor disc memory and additionally hindered the rebuilding of a sufficiently large stellar disc. Looking at the individual cases of slow rotators in our model, we find that on average slow rotators with high fractions of minor mergers went through an episode in their past during which they were classified as fast rotators, while those slow rotators with low minor merger fractions  were always close to the slow rotator regime.   

\section{Fast and Slow Rotator Demographics}\label{demo}
The big advantage of the \atl sample is its completeness, which allows us to directly compare our model population with the observed one.  As we have shown in Fig. \ref{f6} our fiducial model is able to reproduce the  relative fraction of fast and slow rotators as a function of magnitude. We find in addition, that the number density of ETGs with $M_* > 6 \times 10^{9}$ M$_{\odot}$ is $1.6 \times 10^{-3}$ Mpc$^{-3}$, comparing well to the \atl one of $2.2 \times 10^{-3}$ Mpc$^{-3}$ (Paper I). Approximately $ 20 \%$ of these ETGs are slow rotators, in fair agreement with the $14 \pm 2 \%$  found in  the \atl sample (Paper III).  Having established that our model reproduces the basic statistics of the \atl sample  we can now address the question of what the deciding factors are in the history of  an ETG  that make it become a fast or slow rotator within our model. As we argued above we take the disc fraction in ETGs  to first order as a  proxy of $\lambda_R$, thus the balance between disc destruction  and formation plays a crucial role. 

\begin{figure} 
\includegraphics[width=0.45\textwidth]{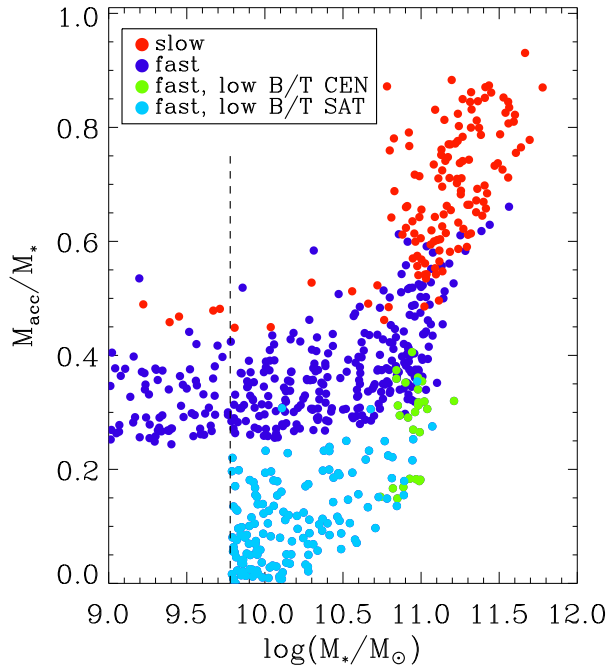}
 \caption{The fraction of accreted stars from satellite galaxies onto ETGs as a function of the present-day galaxy mass.  Massive galaxies accrete the majority of their stellar material. This can be up to $~ 90\%$ of the total mass in the most massive galaxies. The population of fast and slow rotators show a clear distinction in this figure, with the latter (red filled circles) showing $M_{acc}/M_* \geq 0.5$.  Slow rotators show a strong trend with mass due to the cooling shut-off in haloes more massive than $M_{DM,crit}$. Fast rotators with low $B/T$ ratios are dominantly satellite galaxies in high density environments that got stripped of their hot haloes and ceased to form stars (light blue filled circles).  Please note that we do not show satellite galaxies below the \atl survey mass limit of $\sim 6 \times 10^9$ M$_{\odot}$ (dashed line), but show central galaxies all the way down to $10^9$ M$_{\odot}$ to make the trend with mass more visible.     } \label{f8}
\end{figure}

We start by looking at the fraction of accreted stars that end up in ETGs as a function of their stellar mass. By definition these are all stars that a satellite galaxy contributes to the primary galaxy during a merger and directly proportional to the amount of stellar disc that might be destroyed in the primary (see Eq. \ref{eq2}), suggesting that fast rotators should have on average $ M_{acc}/M_* \leq 0.5$ depending on the amount of gas and stars in the progenitor discs.   As demonstrated in Fig. \ref{f8}, the majority of fast rotators has indeed  $M_{acc}/M_* \leq 0.5$  and the amount of accreted material in slow rotators can reach up to $90 \%$ of their total mass. These estimates are in agreement with earlier results of  \citet{ks06}. When dividing the ETG sample into fast and slow rotators a clear separation appears supporting the observational classification into slow and fast rotators. Slow rotators have on average larger fractions of accreted material than fast rotators and show a very steep relation between the accreted mass fraction and their total stellar mass. 

Fast rotators in our model fall into two different classes, those with $0.5 < B/T < f_{fast}  $ $ (\sim 78 \%)$   and those with  $B/T \le  0.5 $ and low gas fractions $( \sim 22 \%)$.  The high $B/T$ population shows a clear separation from the other fast rotators, having higher fractions of accreted material at a given stellar mass. Also clearly visible is a moderate correlation with stellar mass toward higher accretion fractions. The low $B/T$ population on the other hand shows a large scatter and stays generally below accretion fractions of $0.3$. These two populations in terms of their morphology will not necessarily show different properties other than their B/T ratio, and thus would to first order look like members of one homogenous class of fast rotators. However, detailed analyses of their stellar population, in terms of age, metallicity and detailed structure might reveal differences, due to their different formation paths and/or the environment in which they evolved. As we showed earlier, the most dominant formation channel for fast rotators is via the destruction of existing stellar discs in minor mergers resulting in high $B/T$ fast rotators and possibly thickened stellar discs in contrast to the low $B/T$ population of dried-up late-type galaxies with thin stellar discs.

The trend seen in Fig. \ref{f8} can be understood by considering the growth of galaxies. The ratio $M_{acc}/M_*$ increases during dissipationless  mergers and is lowered  during dissipational events, like cooling and associated star formation. The modest stellar mass dependence seen for high $B/T$ fast rotators is mainly driven by the slowing down of cooling in massive haloes. Once host haloes reach masses above $M_{DM,crit}$ cooling completely stops and  galaxies grow mostly via dissipationless mergers \citep{2009MNRAS.397..506K}   causing the strong dependence of $M_{acc}/M_*$ on stellar mass seen in slow rotators. The low $B/T$ fast rotators resemble the flattened fast rotators in the \atl sample and are the result of cold star forming gas running out in either satellites galaxies in dense environments $( \sim 17\%$  of fast rotators)(light blue filled circles in Fig. \ref{f8})  or central galaxies in  haloes $ >    M_{DM,crit}$ that did not experience many mergers during their evolution $( \sim 5 \%)$ (green filled circles). Another $20 \%$ of fast rotators have high $B/T$ ratios and are satellites in dense regions giving a total of $\sim 37 \%$ of fast rotators being satellite galaxies subject to environmental effects (Paper I). The majority of $ \sim 60\%$ however, are central galaxies. In contrast, slow rotators are in over $85 \%$ of the cases central galaxies.
\begin{figure} 
\includegraphics[width=0.45\textwidth]{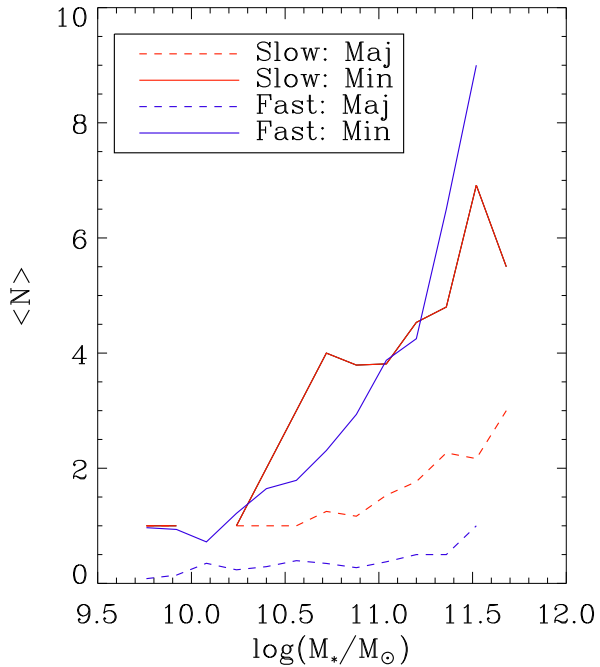}
 \caption{The average number of minor (solid) and major (dashed) mergers that the most massive progenitors of present day ETGs during their evolution experience. The number of mergers rises with the present-day stellar mass of ETGs. Both populations of ETGs show similar numbers of minor mergers as a function of stellar mass. However, slow rotators have on average more major mergers than fast rotators.  The model estimates on the number of mergers is in very good agreement with recent estimates of \citet{2009MNRAS.397..208C}       } \label{f8b}
\end{figure}

Besides differences in the fraction of accreted material it is important to ask whether the accreted material is provided in the same way for fast and slow rotators. In Fig. \ref{f8b} we show the average number of major and minor mergers the most massive progenitor of present-day fast and slow rotators experienced.  Not surprisingly, we find that the number of mergers is an increasing function of stellar mass and that the average number of minor mergers is similar for fast and slow rotators of the same mass. The number of major mergers however, is larger for slow rotators. The average number of major mergers is between 1 and 2, which implies that they encountered at least one major interaction during their formation which would leave its imprint.   

In summary, fast and slow rotators mark the transition point in the ability for gas to cool and form stars in discs. Slow rotators have accreted more stars via minor and major mergers, reducing their disc fraction and hence $\lambda_R$. In addition they have  more major mergers, at least one on average during their evolution that could leave an imprint in their stellar body.

\section{The Redshift Evolution of Fast and Slow Rotators}\label{z}
\begin{figure*}
\includegraphics[width=0.85\textwidth]{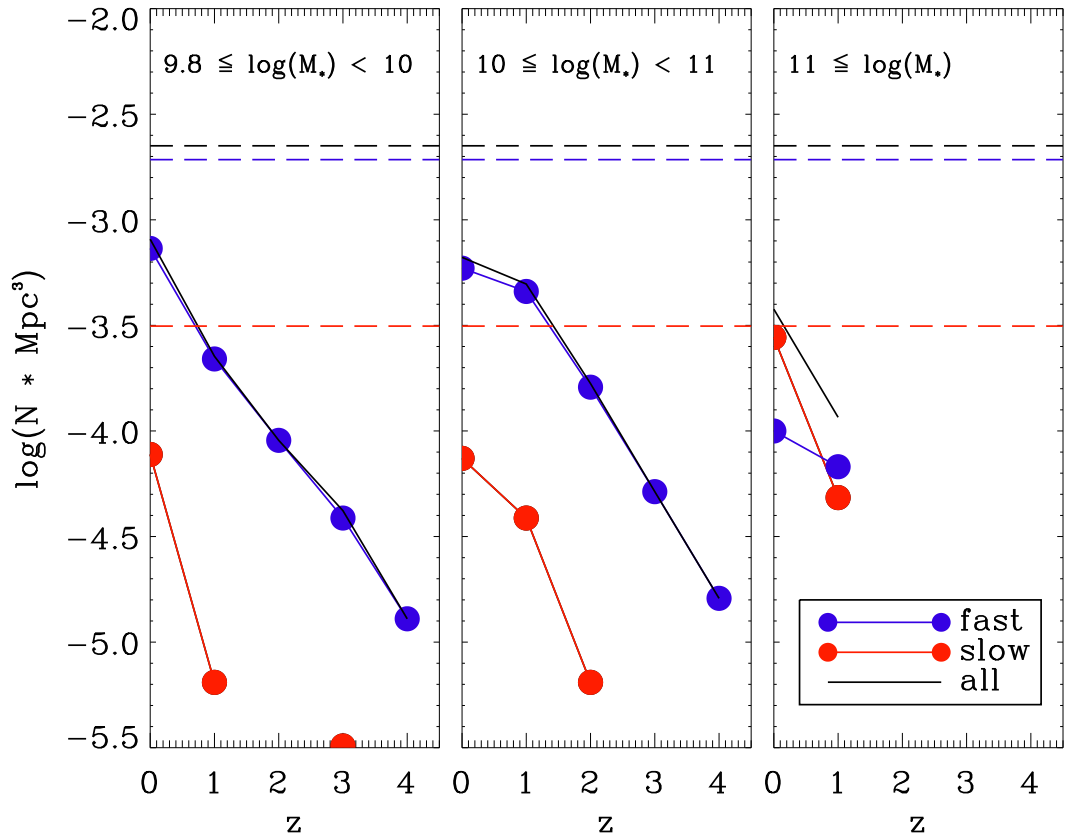}
\caption{ Number density of fast and slow rotators as a function of redshift. We apply  three different stellar mass bins to highlight the mass dependence. The  long dashed lines show the number-density of  slow (red), fast (blue) and all (black) \atl galaxies at $z=0$, respectively. Our model predicts a strong decline of slow rotators  even for the highest mass bin and predicts that fast rotators should be more than an order of magnitude more frequent at $z \geq 2$ for   galaxies with $M_* > 10^{10}$M$_{\odot}$. The high-z number densities of slow rotators are upper limits since they are the results of recent equal mass mergers that under certain merger orbits result in fast rotators (Paper VI). } \label{f10}
\end{figure*}

As we laid out earlier, in our model the ability to cool gas and the fraction of accreted stars via mergers are  important to predict the fraction of fast and slow rotators. The impact of both these physical processes is  changing going to higher redshifts. The merger rate of galaxies peaks around $z \sim  2 -3 $ \citep[e.g.][]{2003AJ....126.1183C}, and the gas accretion rate also does increase towards higher redshifts \citep{2009Natur.457..451D}. In Figure \ref{f10} we present the number density of fast and slow rotating ETGs as a function of redshift for three different  ranges in stellar mass  ($ 9.8 \leq \log M_* < 10$; $ 10 \geq \log M_* < 11 $ and $ \log M_* >11$). The horizontal line shows the number density of early-type galaxies in the  \atl volume. Going to higher redshift the number density drops as expected due to the hierarchical build-up of massive galaxies that takes place, and the fact that a larger fraction of massive galaxies are classified as gas-rich late-type galaxies. Independent of the overall decline, the number density of slow rotators  in all mass bins in general declines stronger with respect to the one of fast rotators. 
Many of the slow rotators at high-z are remnants of very recent equal mass mergers, in contrast to the low-z slow rotators, that have had major mergers in their past but are dominated by minor mergers  during the later stages of their evolution. As shown by Paper VI equal mass mergers that take place under special merger orbits can result in fast rotators. The number densities for slow rotators at high-z that we present are therefore upper limits. Independently, our results predict that beyond $z \sim 2$ it will be hard to find slow rotators and that the population of ETGs at $z \sim 2$ should be dominated by fast rotators, which are more than one magnitude more frequent than slow rotators  at $M_* > 10^{10}$ M$_{\odot}$. Taking the observational results on the evolution of the number-density of massive ETGs at face value \citep[e.g.][]{2009MNRAS.396.1573F} our results suggest that basically all massive slow rotators were fast rotators at some point during their evolution.

\section{Cosmic Variance} \label{var}
The ATLAS$^{3D}-$survey only spans a limited volume of the local universe within a radius of $42$ Mpc. It is thus valid to ask whether cosmic variance has a significant impact on the results of the survey. We try to address this question by using our fiducial model and investigating the effects of cosmic variance using a large scale N-body simulation.

\subsubsection{Variations in the Dark Matter Mass Function}
\begin{figure} 
\includegraphics[width=0.45\textwidth]{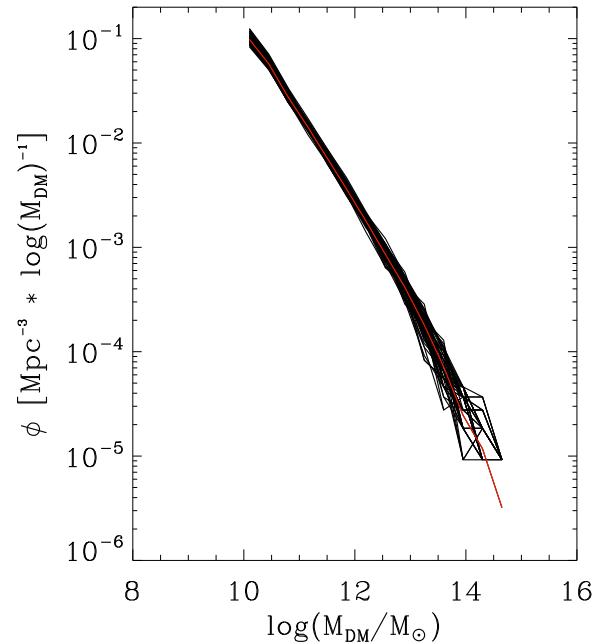}
\caption{ The FOF-mass function of dark matter haloes in a cosmological N-body simulation (red line) of volume $(100$ Mpc$/h)^3$, and in 50 spherical sub-volumes of the size of the \atl volume, placed randomly within the large simulation. The effect of cosmic variance is clearly visible at the massive end of the mass function.   }  \label{f11}
\end{figure}

Depending on the location of a fixed volume within the Universe, the mass function of dark matter host haloes and associated galaxies will change. We estimate this effect, by using a cosmological N-body simulation of a $(100$ Mpc$/h)^3$ volume. The particle resolution of the simulation is $10^9$ M$_{\odot}/h$, and we generate the $z=0$ mass function for haloes using the friends-of-fiends (FOF) algorithm.  Figure \ref{f11} shows the mass function of the whole box (red line) and that of 50 randomly placed spherical sub-volumes of the size of the \atl volume. The mass functions agree well at low masses, but starts deviating at large masses due to cosmic variance. To predict the fraction of fast and slow rotators within each of these sub-volumes we take the FOF-mass function and apply the merger-tree algorithm of \citet{1999MNRAS.305....1S} in combination with our fiducial model. We do not construct the merger trees from the simulation, because we do want to also investigate what the impact of randomly generated merger histories is, choosing  haloes of the same mass  (see $\S$ \ref{var_mer}). 

\begin{figure} 
\includegraphics[width=0.45\textwidth]{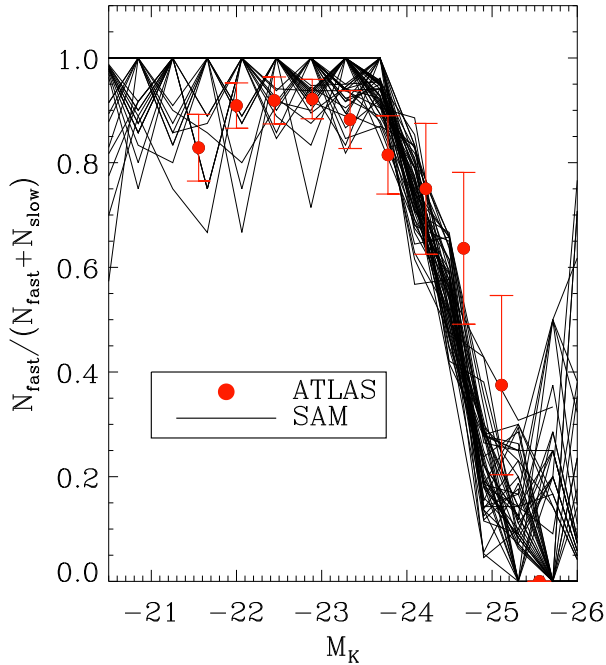}
\caption{ Same as Fig. \ref{f6} for 50 random spherical sub-volumes within a large scale cosmological N-body simulation. The cosmological simulation itself has a volume of $(100$ Mpc$/h)^3$ and each sub-volume has a volume comparable to that covered by the  \atl sample, i.e. $V_{A3D}=  10^5 $ Mpc$^{3}$. Our results show that cosmic variance should not play a large role for the \atl sample at $M_K > -25$. Only the fraction of fast rotators at the very luminous end might be affected by cosmic variance.}  \label{f12}
\end{figure}

The fraction of fast rotators in Fig. \ref{f12} shows fluctuations at $M_K \gtsim -25$, which are well within the error-bars of the observed sample. Only for more luminous galaxies the fraction of fast rotators is heavily affected by cosmic variance. However, the fraction of fast rotators in the highest luminosity bin is on average still below $30 \%$.  In terms of the \atl sample our results suggest that the general trend of a declining fraction of fast rotators as a function of luminosity is robust and not affected by cosmic variance.

\subsubsection{Variations in Merger Histories}\label{var_mer}
\begin{figure} 
\includegraphics[width=0.45\textwidth]{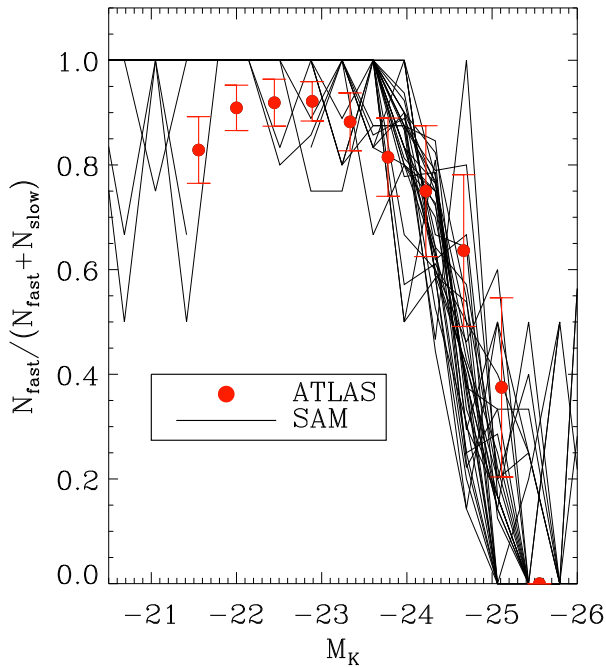}
\caption{ Same as Fig. \ref{f12} but for 30 random merging trees using the same mass function of haloes at $z=0$. Below $M_K \sim -25$ results are robust and within the errors. The fraction of fast rotators at very low and very high luminosity is heavily affected by the individual merging histories of dark matter haloes.  }  \label{f13}
\end{figure}

Besides changes to the dark matter mass function in a given volume the individual merging history of the dark matter haloes might be different as well, and have a significant impact on the population of early-type galaxies. 
We investigate the impact of the merger histories by generating 30 random merging histories for the dark matter mass function drawn from one of the sub-volumes. Again we apply our fiducial model to predict the fraction of fast and slow rotators. The scatter about the median value is well within the error bars of the observations (Fig. \ref{f13}), and large deviations only occur at $M_K \ltsim -25$. The magnitude of the scatter is comparable to the one seen for different mass functions and suggests that part of it might be initially due to the varying merger histories and only to second order by the different number densities of haloes in a given volume. Varying merging histories has most impact at the extreme high-mass tail, where the relatively small number of  major mergers can be strongly influenced by a few more major mergers during the history of a galaxy.

\section{Discussion \& Summary}\label{sum}
In this study we have used a semi-analytic modelling approach that by construction reproduces first order key observations, such as the mass function, the trend in bulge-to-total stellar mass ratio as a function of stellar mass and the gas fraction in late-type galaxies. On top of this model we have made a selection of ETGs that tries to be as close to the \atl  (Paper I) selection as possible. While historically $B/T$ \citep{1999MNRAS.303..188K,1986ApJ...302..564S}  has been the preferred choice, in this study we extended this criterion to take into account the revised definition of ETGs proposed by the \atl survey (Paper I). We introduce the additional condition of low cold gas fractions in disc dominated galaxies to account for the population of flattened gas-poor fast rotating ETGs in the \atl sample. Support for this approach comes from the observed distribution of gas fractions in late-type and early-type galaxies in the \atl parent sample \citep{paolo}, and from the fact that binary mergers in general do not result in fast rotators as flat as observed (Paper VI). Anyway, the fraction of model fast rotators with low $B/T$ and low gas fractions is  only $22 \%$ and is mainly affecting the low mass end, where fast rotators dominate in any case. We thus expect this ETG selection not to change any trend seen between the fraction of fast and slow rotators at high masses.  Based on these ETG selection criteria our model predicts a number density of ETGs in good agreement with the  \atl parent sample. We did not apply any 'fine tuning' to match the number densities of ETGs, and these are just the outcome of the selection procedure that we applied. It is interesting to note, that more than $80 \%$ of ETGs in our sample have $B/T$ values consistent with the classical  selection of E and S0 galaxies based on the Hubble T-type and bulge-to-total light ratios \citep{1986ApJ...302..564S}. Similar agreement has been found between the T-type based morphological definition and the  \atl  ETG selection (Paper I) further supporting that  our model ETG selection is close to the \atl one.    

In our model we grow bulges only via mergers and we neglect internal processes like disc instabilities that have been suggested as additional ways to grow bulges \citep[e.g.][]{2009MNRAS.396.1972P,2009ApJ...703..785D}. These processes are most efficient in gas-rich massive discs and would result in an increased fraction of fast rotators with respect to our fiducial model with a range in $B/T$ ratios depending on when the disc becomes stable. In such models clumps form in the disc and migrate due to dynamical friction into a central bulge. It is however, not clear at this point whether such clumps survive long enough to reach the centre or get disrupted by feedback from supernovae \citep{2010arXiv1011.0433G}. 

Another important ingredient in our model is the ability to cool gas and re-grow discs. While SAMs in general reproduce the average cooling behaviour of gas in SPH simulations there are difference when comparing on an object-to-object basis \citep{2010MNRAS.406..729S}. Even between individual SAMs deviations in the calculated cooling rates exist \citep{2010MNRAS.406.1533D}. In general, the fraction of massive slow rotators is not affected by changes in the cooling rate, due to the constraint that the galaxy mass function needs to be reproduced. This is usually achieved by  feedback implementations in the SAMs, which regulate the cooling of gas and the overall star formation. SPH simulations show that feedback plays indeed a key role in regulating gas accretion rates and star formation \citep[e.g.][]{2010arXiv1011.2491V,2010MNRAS.402.1536S}. In our model we 'tune' our feedback efficiency and hence our cooling rate in such a way that we reproduce by construction the following observables: the stellar mass function, the gas fraction in late-type galaxies and the cooling rate in galaxies at high z. Using this approach we try to minimise the freedom in our cooling rates. 

Besides cooling of gas, mass loss from already existing stars can provide significant material for the growth of stellar discs \citep{2010ApJ...714L.275M,2010MNRAS.tmp.1527A,ts}. We do not take this into account in our model, but note here that such mass loss is likely to decrease  the $B/T$ ratio of galaxies. \citet{2010ApJ...714L.275M} find that the Hubble T-type of a galaxy can change via this process by 2 to 3 units, affecting the fraction of fast rotators.

The merger rate of galaxies is  a key aspect in our modelling approach. Being able to reproduce the stellar mass function, we populate dark matter haloes on average with galaxies of the right mass. Thus when dark matter haloes merge we expect to merge the 'right' galaxies as well. In fact, the derived  merger rates and average number of mergers from our model are in very good agreement  with observations \citep{2009ApJ...697.1971J,2009MNRAS.397..208C}.

Within the ETGs selection we distinguish between fast and slow rotators based on the fraction of stellar disc still present in the main body. The main reasoning behind this approach is the observational fact that many fast rotators show regular rotation patterns and signs of bars (Paper II), which indicate the presence of a disc-like component (Paper VII). In addition disc galaxies show values of $\lambda_R > 0.5$ similar to fast rotating ETGs in the \atl sample (Paper VI).  In Paper VI the results of binary mergers have been extensively compared to the structure of galaxies in the \atl sample showing that binary disc mergers reproduce best the properties of fast rotators with $\lambda_R \sim 0.25-0.5$, while fast rotators with $ \lambda_R < 0.25$ are mostly reproduced in re-mergers of disc merger remnants, and that the majority of slow rotators cannot have formed from a binary major merger or re-merger. They in general have a much more complex formation history that will be investigated in \citet{naab} and \cite{bois}. Paper VI shows that major mergers efficiently reduce $\lambda_R$ in the central regions. Minor mergers on the contrary are expected to reduce $\lambda_R$ in the outer regions. These results suggest that within the fast rotator population a sequence of mergers will preferentially lead to a reduction of $\lambda_R$. The way this can be naturally achieved is by the  destruction of progenitor disc components via violent relaxation during  mergers. However, discs can re-build by gas accretion and star formation increasing $\lambda_R$, and  it is therefore important to model the complete merging history of a present-day ETG to decide on its status in terms of fast or slow rotation. We here do not attempt to model the detailed structures of fast and slow rotating ETGs but try to give answers on the average formation path of these  galaxies. To achieve this we introduce a free model parameter, the minimum disc fraction required in fast rotators to divide our model ETG sample into fast and slow rotators. This parameter has been set to $f_{fast}=0.1$ by matching the fraction of fast and slow rotators as a function of magnitude  and their number density at $z=0$, and serves the dual purpose of a free model parameter as well as  a prediction of the expected disc fraction in fast rotators that can be compared to observations, once proper bulge-disc decompositions of \atl galaxies have been performed (\atl in prep.). Effectively our model assumes that $\lambda_R$ is to first order a proxy for the disc fraction in ETGs.  

The modelled fast and slow rotators show clear distinct growth histories, supporting the observationally motivated separation into two classes. The two main differences are that slow rotators grow by more accretion of stellar mass from satellites than fast rotators of the same mass, and that they have on average more major mergers.   

The fraction of accreted material depends strongly on the competing effects of disc growth in galaxies and their destruction during mergers. While fast rotators continue accreting small levels of gas and forming stars in their recent history, consistent with observed star formation in  ETGs \citep{2009MNRAS.396..818S}, slow rotators completely stop accreting gas, and only grow via mergers. The main mechanism causing cooling to stop in high mass galaxies is still debated and could have several different origins \citep[see e.g.][]{2006MNRAS.365...11C,2006MNRAS.370.1651C,2008ApJ...680...54K,2009ApJ...697L..38J}. We here adopt an empirical  sharp transition in the cooling behaviour of gas based on a critical  dark halo mass. It is likely that such a transition should in reality be smooth based on the underlying  physics.  Introducing a scatter in the critical halo mass, to model a smooth transition, is not changing the general growth properties of fast and slow rotators though. It is therefore safe to assume that whatever physical process causes cooling to stop in massive haloes, as long as it scales with galaxy mass, respectively halo mass, it will not change the growth history of fast and slow rotators.  Completely neglecting any cooling shut-off results in too many massive galaxies \citep{2006MNRAS.365...11C}, and in particular too many massive  fast rotators. 

The higher accretion fraction in slow rotators has another direct implication for the hot halo gas. The potential energy of infalling satellite galaxies is an effective source of gravitational heating of the hot halo gas \citep{2008ApJ...680...54K,2009ApJ...697L..38J}. Either via dynamical friction heating \citep{2004MNRAS.354..169E}  or shocks large parts of the potential energy  transfers to the halo gas. Our model thus naturally predicts to find preferentially  hot X-ray haloes around slow rotators, and not fast rotators of the same mass \citep{2010MNRAS.402.2187S}, suggesting that X-ray haloes are a consequence of the mass assembly of an ETG.

Another important consequence of the higher accretion fractions in slow rotators is that even though they have the same average number of minor mergers as fast rotators of the same mass, the masses of the satellites are higher and the relative contribution towards lowering $\lambda_R$ is much more important.  Higher accretion fraction means  higher disc destruction fraction  through violent relaxation during the evolution of a galaxy, and thus lower $\lambda_R$. We find on average less than 10 minor mergers for the most massive progenitors of fast and slow rotators (Fig. \ref{f10}), which allows the mass distribution of merging satellites to be different for  fast and slow rotators.

The average cold gas fraction of satellite galaxies merging with progenitors of present-day slow rotators is lower than that of fast rotators. This bias is a direct consequence of the fact that the infalling galaxies in slow rotators are more massive and thus have lower gas fractions (see Fig. \ref{f5}), which hinders additionally the rebuilding of discs.    

The second main difference in the growth history of fast and slow rotators is the number of major mergers. Slow rotators have on average more than twice as many major mergers than fast rotators. However, even the most massive ones do not have more than 3 major mergers on average during their evolution. We find that most of these major mergers happen at early times between gas-rich progenitors comparable to those simulated in Paper VI and \citet{2010ApJ...722.1666W}. Such gas rich major mergers produce  kinematically distinct cores (KDC) similar to ones observed in present-day slow rotators (Paper VI). However, the remnants from  major merger simulations tend to be not round enough compared to observations. Based on the  merging history in our SAM we find that many minor mergers follow such a major merger event. Minor mergers from random directions \citep{2006A&A...445..403K}, if not too compact, get stripped of their stellar material in the outer parts of the host galaxy, providing an envelope  of stars that can make too flat major merger remnants with KDCs become more round. Further indirect evidence for the importance of minor mergers for the growth of a host galaxy particularly in its outer layers comes from the observed size-evolution of ETGs \citep[e.g.][]{2006ApJ...648L..21K,2007MNRAS.382..109T,2009ApJ...697.1290B,2009ApJ...699L.178N,2010MNRAS.401.1099H}. As shown in Paper VI KDCs are very fragile during major mergers. Their observed presence in slow rotators prompts the question how they can survive till today. Analysing    
the growth history of slow rotators in our model, we find that in most of the cases the last gas-rich major merger was not very recent $( z > 1.5 )$  (see however, \citet{pap9} for the case of a remnant from a possible recent gas-rich major merger), and that the number of major mergers is just not very high (see Fig. \ref {f8b}), both favouring the survival of KDCs in slow rotators. In contrast fast rotators have on average less than one major merger in their history, making them unlikely candidates to host large-scale KDCs. 

Within our model we predict a strong evolution of the slow rotator fraction toward low z. At  $ z \sim 2$ the number density of massive  ($M_* > 10^{10}$ M$_{\odot}$) fast rotators is more than one order of magnitude larger than the slow rotators one. This is mainly due to high cooling rates and gas fractions in mergers promoting stellar disc building over destruction. The evolution toward low redshifts is driven by the inability to efficiently re-build stellar disc in massive host galaxies, which is another representation of down-sizing in star formation.   

\section{Conclusions}\label{con}
In this paper we made use of the completeness of the \atl sample to investigate the origin and formation history of fast and slow rotator ETGs within a self-consistent cosmological framework using the semi-analytical modelling approach. We  here present a model in which  the difference between fast and slow rotator ETGs is purely based on the stellar disc fraction found in them, and predict that fast rotators have disc fractions  $M_{disc}/M_* > 0.1$. We find that  slow rotators within an evolving universe mark the transition in the ability of galaxies to cool  gas and to rebuild stellar discs.  We find a clear separation in the growth history of fast and slow rotator ETGs, supporting the observationally motivated distinction. In particular the accreted fraction of stars shows a clear distinction between fast and slow rotators, with the latter having between $50 \% - 90 \%$ of their stellar mass  accreted from satellites while the former has $ < 50 \%$ accreted.

Although we find a clear separation into fast and slow rotators, we also find that ETGs can switch their state between fast and slow rotator and vice versa  based on stellar disc growth or destruction by mergers, suggesting that fast and slow rotator ETGs are transient. These changes however, occur predominantly at higher redshifts when cooling is more efficient and mergers more frequent. The fraction of slow rotators,  shows a continued increases with time due to the conversion of fast rotators and the inability of gas to cool and convert slow rotators back to fast rotators. Massive present-day slow rotators therefore, can be viewed  as the final stage in the evolution of  ETGs. 

Future high redshift observations of ETGs will be able to reveal any possible evolution in the fraction of fast and slow rotators and test the presented model  further.  

\section*{Acknowledgments}
We would like to thank the referee for helpful comments.
MC acknowledges support from a STFC Advanced Fellowship PP/D005574/1 and a Royal Society University Research Fellowship. This work was supported by the rolling grants ÔAstrophysics at OxfordÕ PP/E001114/1 and ST/H002456/1 and visitors grants PPA/V/S/2002/00553, PP/E001564/1 and ST/H504862/1 from the UK Research Councils. RLD acknowledges travel and computer grants from Christ Church, Oxford and support from the Royal Society in the form of a Wolfson Merit Award 502011.K502/jd. RLD also acknowledges the support of the ESO Visitor Programme which funded a 3 month stay in 2010. SK acknowledges sup- port from the the Royal Society Joint Projects Grant JP0869822. RMcD is supported by the Gemini Observatory, which is operated by the Association of Universities for Research in Astronomy, Inc., on behalf of the international Gemini partnership of Argentina, Australia, Brazil, Canada, Chile, the United Kingdom, and the United States of America. TN, SK and MBois acknowledge support from the DFG Cluster of Excellence ÔOrigin and Structure of the UniverseÕ. MS acknowledges support from a STFC Advanced Fellowship ST/F009186/1. NS and TD acknowledge support from an STFC studentship. The authors acknowledge financial support from ESO. The SAURON observations were obtained at the William Herschel Telescope, operated by the Isaac Newton Group in the Spanish Observatorio del Roque de los Muchachos of the Instituto de Astrofisica de Canarias. This research has made use of the NASA/IPAC Extragalactic Database (NED) which is operated by
the Jet Propulsion Laboratory, California Institute of Technology, under contract with the National Aeronautics and Space Administration. We acknowledge the usage of the HyperLeda database (http://leda.univ-lyon1.fr). Funding for the SDSS and SDSS-II was provided by the Alfred P. Sloan Foundation, the Participating Institutions, the National Science Foundation, the U.S. Department of Energy, the National Aeronautics and Space Administration, the Japanese Monbukagakusho, the Max Planck Society, and the Higher Education Funding Council for England. The SDSS was managed by the Astrophysical Research Consortium for the Participating Institutions. This publication makes use of data products from the Two Micron All Sky Survey, which is a joint project of the University of Massachusetts and the Infrared Processing and Analysis Center/California Institute of Technology, funded by the National Aeronautics and Space Administration and the National Science Foundation.



\label{lastpage}

\end{document}